\documentclass[12pt]{iopart}
\usepackage{epsfig}
\usepackage{colordvi}
\newcommand{\gsim}      {\mbox{\raisebox{-0.4ex}{$\;\stackrel{>}{\scriptstyle \sim}\;$}}}

\newcommand{\ptr}      {\mbox{$p_{\rm T}$}}
\newcommand{\ptds}      {\mbox{$p_{\rm T}(D^*)$}}
\newcommand{\etads}      {\mbox{$\eta (D^*)$}}

\newcommand{\dsp}       {\mbox{$ D^{*+}$}}

\newcommand{\ds}       {\mbox{$ D^{*}$}}
\newcommand{\dz}       {\mbox{$ D^{0}$}}

\newcommand{\xd}      {\mbox{$x(D^*)$}}

\newcommand{\ftwoccb}   {\mbox{$F_2^{c\bar{c}}$}}
\newcommand{\ftwo}      {\mbox{$F_2$}}
\newcommand{\qsq}      {\mbox{$Q^2$}}

\newcommand{\AmS}{{\protect\the\textfont2
  A\kern-.1667em\lower.5ex\hbox{M}\kern-.125emS}}

\begin{document}
\hyphenation{author another created financial paper re-commend-ed}

\title{Open charm production at HERA}

\author{Ignacio Redondo
.\\
On behalf of the H1 and ZEUS Collaborations.}
\address{Dept. de F\'{\i}sica Te\'orica,
        Universidad Aut\'onoma de Madrid,
       Cantoblanco, Madrid 28049, SPAIN. e-mail: redondo@mail.desy.de}

\begin{abstract}
Measurements of charmed  particle cross sections 
at HERA  in the photoproduction and deep inelastic  regimes are reviewed. The   status of 
the comparison  with perturbative  QCD calculations is discussed. 
\end{abstract}


\vspace{-0.8cm}

\section{Introduction}

\vspace{-0.2cm}

\begin{figure*}[htb]
\begin{minipage}{0.34\linewidth}
\epsfig{figure=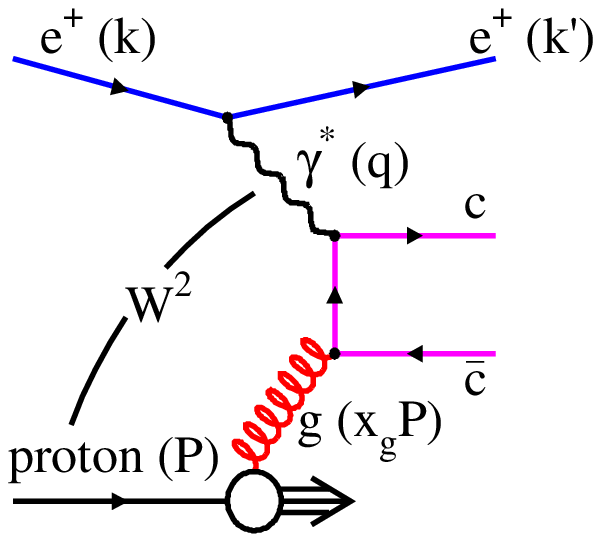 ,width=1.\linewidth}
\begin{picture}(1,1)(0,0)
\put(-10,110){(a)}
\put(250,110){(b)}
\put(400,100){(c)}
\end{picture}
\end{minipage}
\hfill
\begin{minipage}{0.3\linewidth}
\epsfig{figure=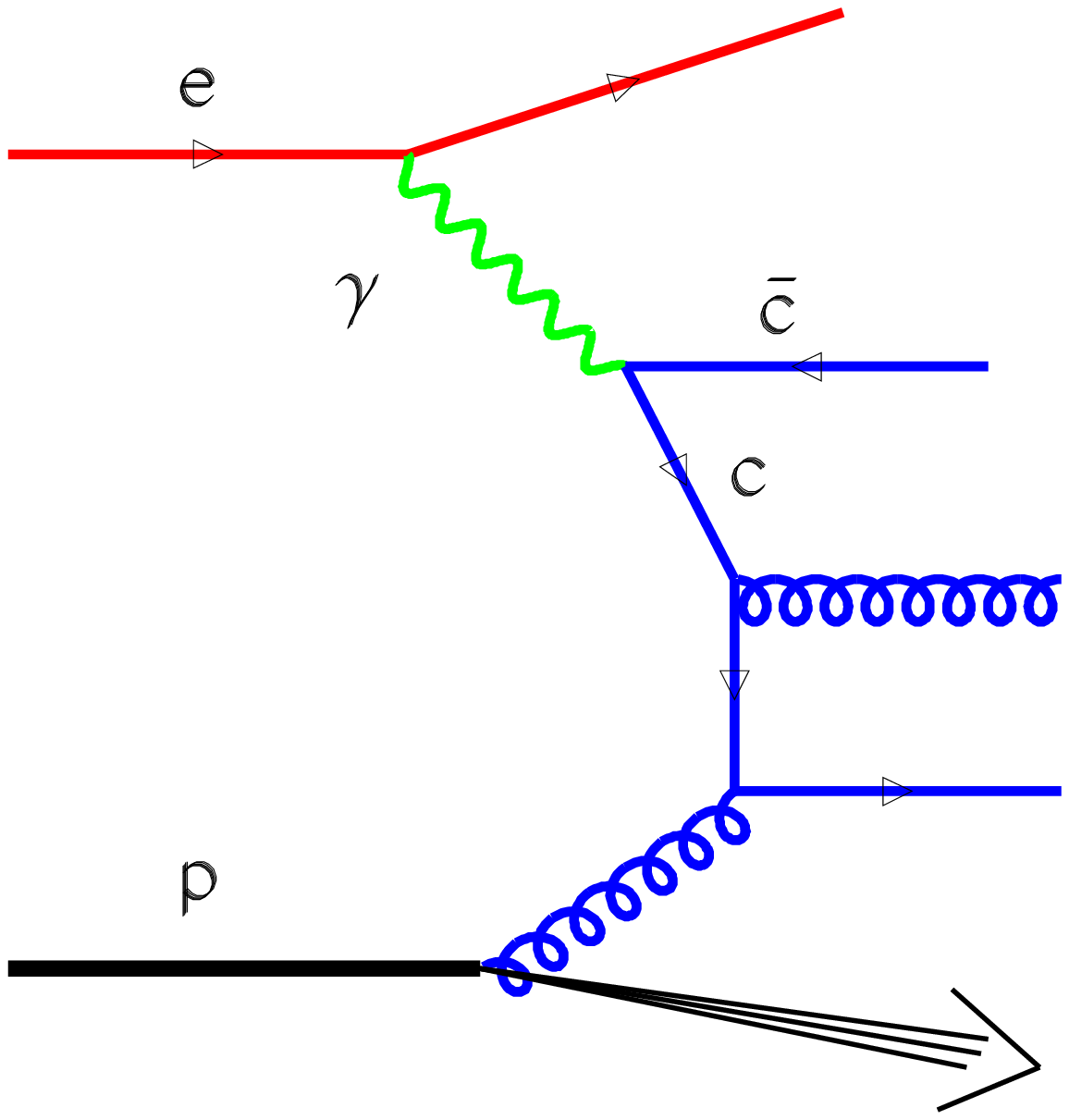 ,width=0.8\linewidth,clip=0}
\end{minipage}
\hfill 
\begin{minipage}{0.3\linewidth}
\epsfig{figure=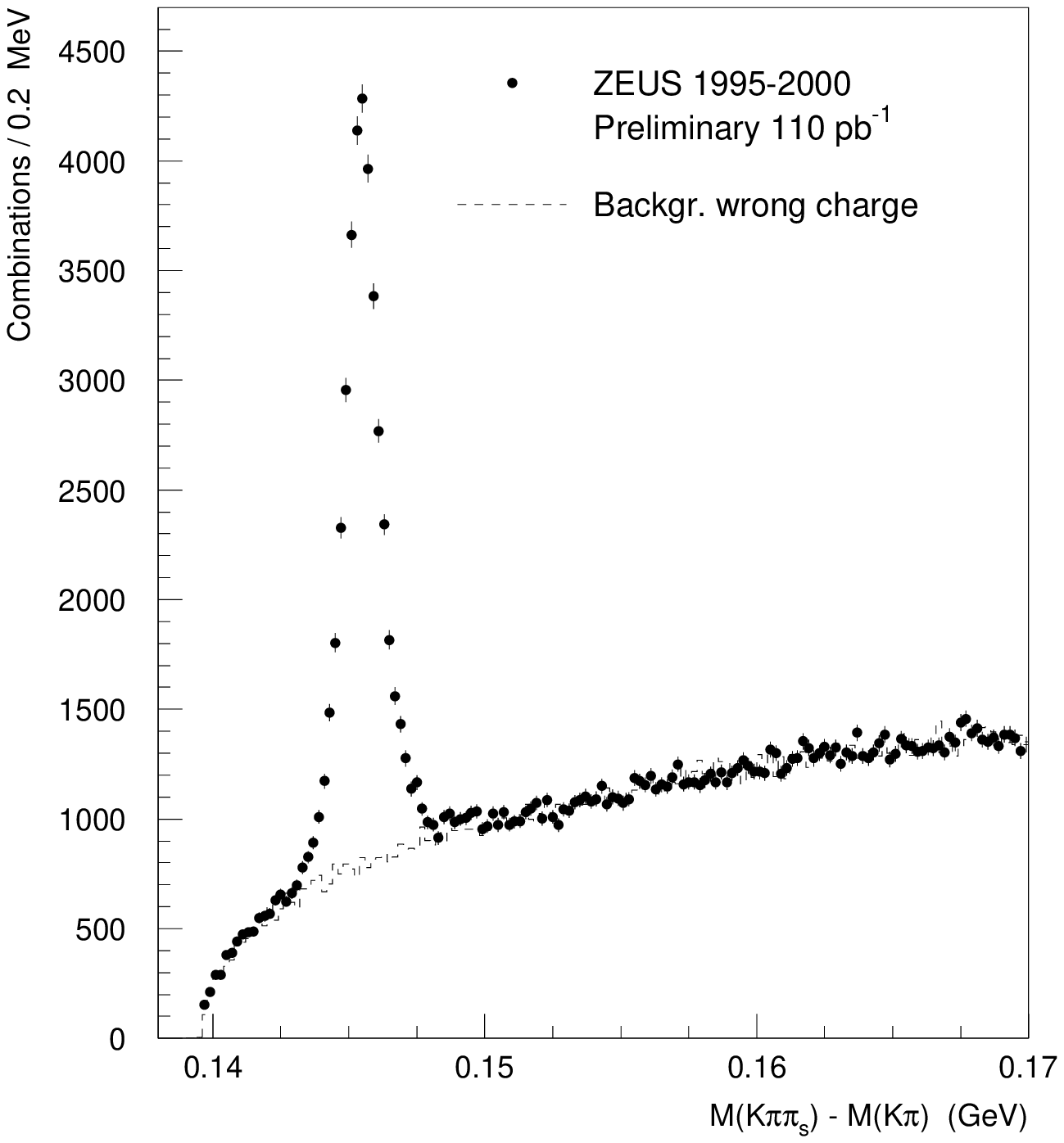 ,width=1.2\linewidth}
\end{minipage}
\caption{ 
 (a) Sketch of the PGF process  and  (b) charm excitation diagram. (c)  \ds\ peak from ~\cite{spectroscopy}.
}\label{f:bgf}
\end{figure*}

 Open charm  production at HERA has developed as a precision testing ground 
for  fixed order (FO) NLO pQCD calculations as well as for other QCD 
inspired models. The bulk of the available data is  on inclusive charmed 
particle production cross sections for several species. I will focus
on reviewing the status of the comparison of the data with FO NLO pQCD
calculations trying to identify the open issues for HERA Run II.
  The description of the kinematics 
for these observables requires two variables for the charmed 
particle momentum (an integration on the azimuthal angle of the
 charmed particle is generally assumed)  and   
one ($W$)  or two ($Q^2$ and $y$, for instance)  additional variables needed 
for the description  of the inclusive  photoproduction 
(PHP)  or deep inelastic (DIS) cross section, respectively.
The  pseudorapidity, $\eta=-log(tan(\theta/2))$, where $\theta$ is the polar angle 
with respect to the proton ($p$) direction, and the transverse momentum with respect to the beam line, \ptr,
 are typically chosen as the  charmed particle variables.
 Open charm production is intimately related to the 
gluon parton density in the $p$ because the Photon-Gluon Fusion (PGF) process, 
sketched in Figure~\ref{f:bgf} (a), is expected to be one of the  dominant production mechanisms. 
 Since the invariant mass of the partonic final state, $M_{hard}^2$, is significant 
due to the presence  of two heavy quarks ($M_{hard}^2>4m_c^2$), the kinematics are such that
the fraction of the $p$ momentum carried by the gluon, $x_{g}=x_{Bj}(M_{hard}^2+Q^2)/Q^2$ rather
 than equal to $x_{Bj}$,  as in the quark parton model (QPM) type of process. 
The sensitivity to the gluon density, $f_g$, and to   $\alpha_s$ can be read from the
factorization formula for the PGF process written in equation (1), where  $\hat{\sigma}$
is the calculable partonic cross section.
\begin{equation}
\hspace{-2.8cm}
\centerline{$\sigma^{c\bar{c}}(x_{Bj},Q^2)\propto  e_c^2 \alpha_s(\mu_r)
\int^1_{x_g^{min}} \frac{dx_g}{x_g} f_g(x_g,\mu_f) \hat{\sigma}(x_g,\mu_f,\mu_r)~~~;~~~~x_g^{min}=x_{Bj}\frac{4m_c^2+Q^2}{Q^2}
$}
\end{equation}
Higher order corrections and the fact that the cross section is measured 
in restricted regions of the charmed particle phase space complicate this 
 simple picture.

 HERA Run I  is already finished but the full data samples 
are just beginning to be exploited.  H1 and ZEUS have   
collected $\gsim$ 100 pb$^{-1}$. This luminosity allows for  the
selection of large charm samples. This is illustrated 
in Figure~\ref{f:bgf} (c), from \cite{spectroscopy}, 
where $\sim$ 27000 \ds\ mesons were reconstructed via the widely used
 K2$\pi$, $D^{\ast +}\rightarrow 
(D^0 \rightarrow K^{-} \pi^+) \pi^+ $(+\,c.c.),  decay channel~\cite{h194,h196,h197,dsdijet,zds44,zds9697,ds318}. 
This decay mode has a low combinatorial background and a sizable branching ratio (0.0262$\pm$0.0010).   
\ds\ mesons have been tagged also via the  K4$\pi$, $D^{\ast +}\rightarrow 
(D^0 \rightarrow K^{-} \pi^+) \pi^+ \pi^+ \pi^-$(+\,c.c.),  mode~\cite{dsdijet,zds9697} which has
a factor of two larger branching ratio (0.051$\pm$0.003) but a larger 
combinatorial background.  There are also data on $D_s$~\cite{dsubs} and 
\dz\ mesons~\cite{h194,zeusdzero}.
Another window to study open charm is the identification of $e^-$ coming 
from semileptonic decays~\cite{semileptonic}. This approach profits from the large branching
fraction $f(\bar{c} \rightarrow e^-)=0.095\pm0.009$.

\vspace{-0.8cm}

\vspace{-0.6cm}
 
\section{Open charm photoproduction}
\vspace{-0.2cm}

Real photon ($\gamma$)-$p$ interactions have a hadronic component in which the $\gamma$
fluctuates into a hadronic final state (resolved). Thus, factorization has to be applied 
also for the $\gamma$ and pdf's have to be measured.
 In addition to the direct PGF component, a large  resolved contribution   comes from 
the `` charm excitation'' process, Figure~\ref{f:bgf} (b).

\begin{figure*}[htb]
\begin{minipage}{0.55\linewidth}
\epsfig{figure=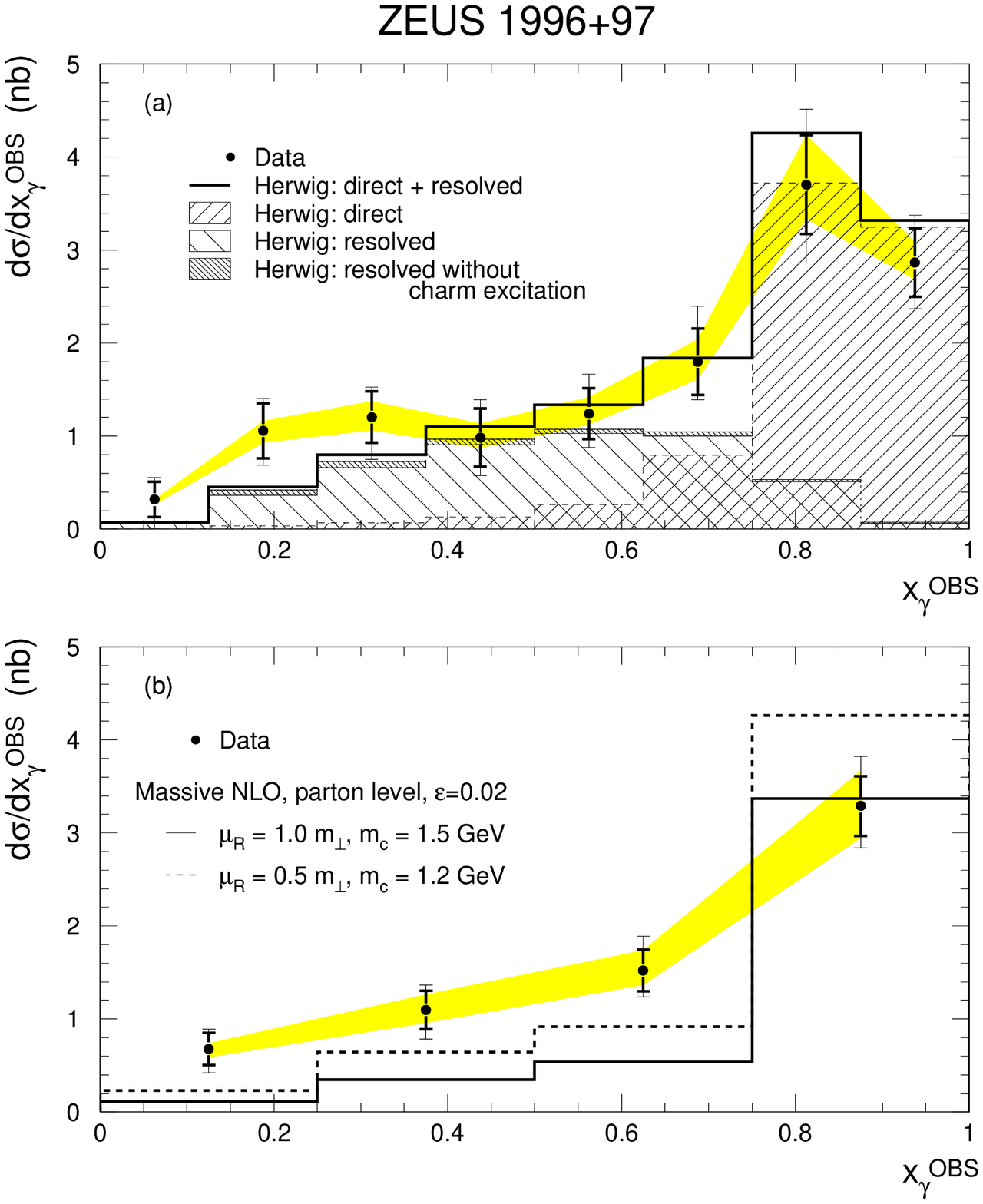 ,width=1.\linewidth}
\end{minipage}
\begin{minipage}{0.44\linewidth}
 \epsfig{figure=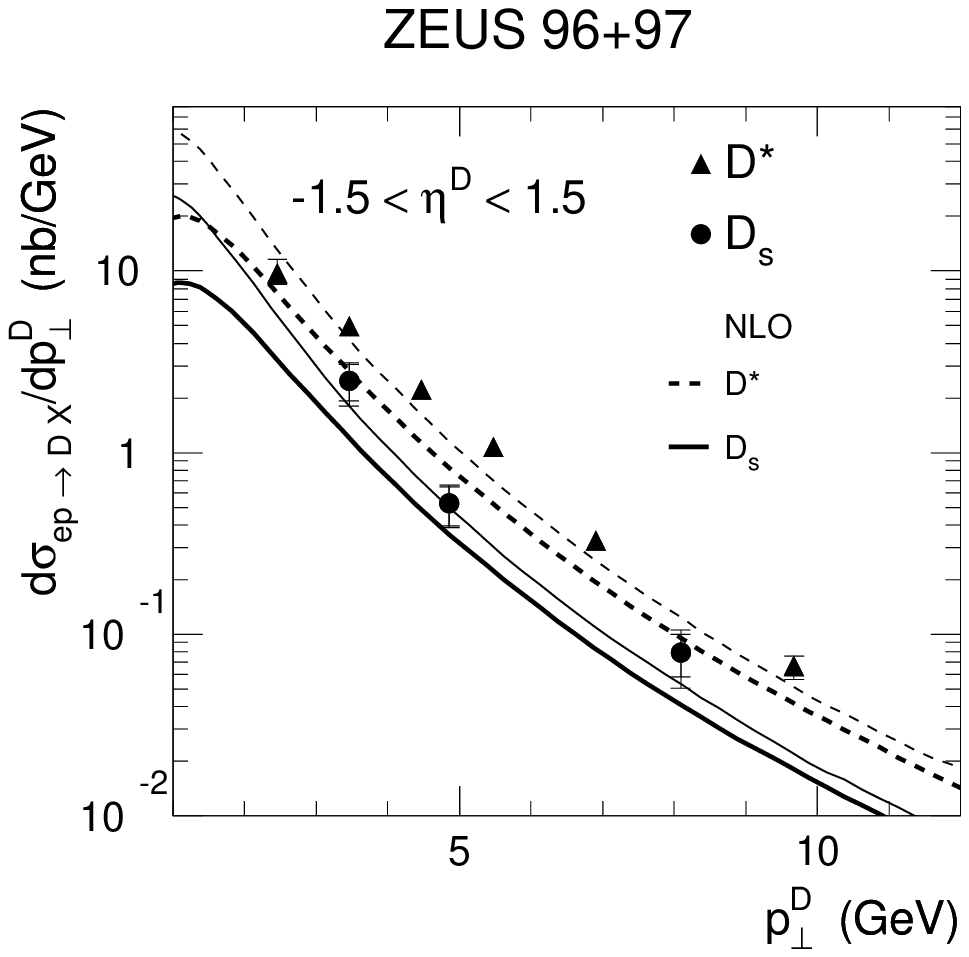 ,width=0.8\linewidth}
 \epsfig{figure=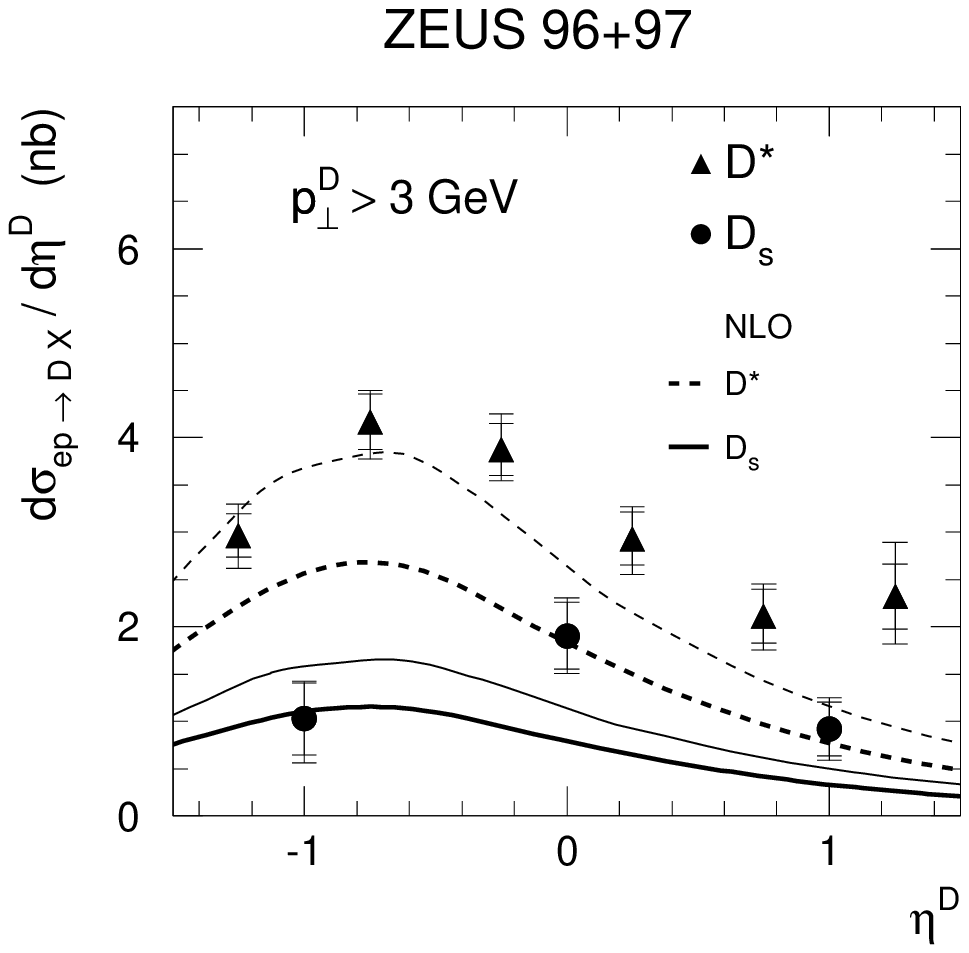 ,width=0.8\linewidth}
\end{minipage}
\caption{ 
Untagged PHP \ds\ dijet cross section differential in $x_{\gamma}^{OBS}$  compared to HERWIG (a) 
and FMNR (b). Inclusive $D$ meson differential  cross sections $d\sigma/dp_{\perp}^D$ (c )
and $d\sigma/d\eta^D$ (d) where $D$ stands 
for $D_s$ (dots) and \ds\ (triangles).  Thick (thin) curves in (b) (c) (d) are FMNR predictions with 
parameters set to $m_c=1.5(1.2)$~GeV, $\mu_R=\sqrt{m_c^2+p_T^2}(0.5\sqrt{m_c^2+p_T^2})$
and $\mu_F=2(4)\mu_R$. 
}\label{f:dsdijet}
\begin{picture}(1,1)(0,0)
\put(380,350){\begin{footnotesize}(c)\end{footnotesize}}
\put(380,195){\begin{footnotesize}(d)\end{footnotesize}}
\end{picture}
\end{figure*}

 The ZEUS Collaboration~\cite{dsdijet} has measured the $x_{\gamma}^{OBS} = \Sigma_{jets} E_T e^{-\eta}/(2yE_e)$ distribution using \ds\ dijets. 
 The separation between resolved and direct is unique
at LO, becoming  scheme dependent at NLO.
 The comparison of the ZEUS data with the HERWIG LO
predictions, Figure~\ref{f:dsdijet} (a), yields $\sim 40\%$ resolved component in the kinematic region 
of the measurement, dominated by charm excitation.
 Figure ~\ref{f:dsdijet} (b) displays the comparison of the dijet data with a massive NLO pQCD 
 prediction produced by the FMNR program~\cite{FRIXIONE}, 
 which underestimates the low $x_{\gamma}^{OBS}$ values (resolved).
 In the  massive approach only light quarks and the gluon pdf's are present in the $p$ 
and the $\gamma$.  Another approach is to neglect the mass of the 
charm quark~\cite{massless} and treat it as a massless quark. 
The resummation of logs in $\ptr/m_c$ results in a charm pdf.
 The main questions in open charm photoproduction are what is the relative weight  of the
 available production mechanisms and what is the more appropriate  model 
in the HERA range, massive or massless. However two caveats limit the precision of the test:
It is known that the $\gamma$ pdf's, obtained from fits to $F_2^{\gamma}$, are poorly known
in the HERA region  and a new iteration including PHP HERA data would be very welcomed. 
 Moreover,  pQCD cannot predict the fragmentation from a charm quark into a hadron. 
 Thus hadron fractions
and fragmentation functions are taken from $e^+e^-$ processes.  
The Peterson fragmentation
 function~\cite{peterson}, $P(z) \propto z^{-1} [1-1/z-\epsilon/(1-z)]^{-2}$  is 
convoluted with the charm quark distribution produced by the partonic matrix element, 
scaling the momentum of the charm quark to that of the \ds\ meson by $z$. Note that
this model is not Lorentz invariant because of the masses involved.
 $f(c \rightarrow \dsp)=0.235 \pm 0.007 \pm 0.007$~\cite{GLADILIN} and $\epsilon=0.035$~\cite{epsilon0.035} are typically used for  \ds\ cross section  predictions.

\vspace{-0.5cm}

\vspace{-0.5cm}

ZEUS has measured \ds~\cite{dsdijet} and $D_s$~\cite{dsubs} untagged PHP, where the 
scattered lepton escapes  undetected through the beam hole, using 38 pb$^{-1}$ of $\sqrt{s}=300$ 
GeV data.  The kinematic region of the untagged  measurements is:
$130<W<280~$GeV,  $Q^2<1~$GeV$^2$ ~($Q^2_{\rm median}\approx 3 \cdot 10^{-4}$~GeV$^2$ ).
The $D_s$ mesons are reconstructed using the 
$D_s^+ \rightarrow \phi \pi^+ \rightarrow (K^+ K^-)\pi^+(+c.c.)$ channel. 
Experimental restrictions result in the phase space limitation
$3<p_T(D_s)<12~$GeV and  $\mid \eta(D_s) \mid< 1.5$.

The fraction of the $D_s$ to the \ds\ cross sections in the above phase space region 
 is measured to be
 $\frac{\sigma(D_s)}{\sigma(D^{\ast})}=0.41\pm 0.07(stat.)^{+ 0.03}_{-0.05} (syst.)\pm 0.10 (br.)$.
 This quantity  is sensitive  within the string model to the strange 
suppression parameter, $\gamma_s$.   $\gamma_s$  $=0.27\pm0.05\pm0.07(br.)$ is obtained with
 the help of JETSET, being  in agreement with $e^+e^-$  and 
therefore supporting  the universality of $c$ fragmentation found in  the earlier  H1 publication~\cite{h194},
and the more recent ZEUS results~\cite{zeusdzero,ds1budapest}.

Figures~\ref{f:dsdijet} (c) and (d) display differential cross sections of 
$D_s$ and \ds\ in untagged PHP. Most of the branching ratio error, which dominates 
 the $\gamma_s$ measurement, can be neglected in the comparison with the calculation 
 since it scales the data and the prediction in the same way.
 The curves are the FO NLO pQCD prediction from FMNR 
using MRSG (p, $\alpha_s$=0.114) and GRV-G HO ($\gamma$, $\alpha_s$=0.111) pdf's and 
the Peterson fragmentation parameter $\epsilon$=0.035. 
Normalization uncertainties coming from the luminosity measurement (1.7\%) and
 the hadronization  fractions (3\% for \ds and 9\% for $D_s$) are not plotted.
The $D_s$ data support the same conclusions of the more accurate \ds\ data, namely:
\begin{itemize}
\item The normalization of the  theory is too low.
\item The shape of the $\eta$ distribution is not well described.
\end{itemize}

\begin{figure}[htb]
\begin{minipage}{0.7\linewidth}
\begin{center} 
\epsfig{figure=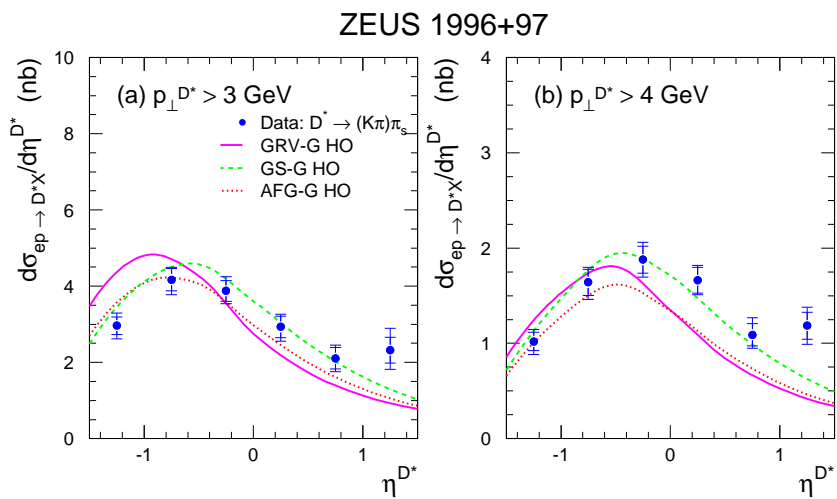 ,width=1.\linewidth}
\end{center}
\end{minipage}
\hfill
\begin{minipage}{0.29\linewidth}
\caption{ Untagged PHP differential cross sections $d\sigma/d\eta^{\ds}$ compared to massless 
NLO calculations (see text)  using different $\gamma$ pdf's.
}\label{f:kniel}
\end{minipage}
\end{figure}

 Massless predictions  give a better description of the data both in shape and normalization 
as  can be seen in  Figure~\ref{f:kniel}. 
The parameters for  the calculation are pdf CTEQ4M (p, $\alpha_s=0.116$),
   $\epsilon=0.1$, $\mu_R=\sqrt{m_c^2+p_T^2}~ ;~ m_c=1.5~$GeV$~ ;~ \mu_F=2\mu_R$. The difference 
 between the curves is the $\gamma$ pdf.
 Surprisingly enough, the predictions using  GS-G HO, with a c pdf equal to the u pdf,
 are the closest to the data.
 The better agreement of the massless calculations is not expected since 
the resummation effects are small for \ptds$<$10 GeV, i.e. in the region of 
the measurement~\cite{frixionedis01}.

 ZEUS has also measured tagged PHP~\cite{zds44} yielding a similar level of description by FMNR.
These data are better described by the BKL model~\cite{bkl},  which fits the untagged sample.  

\begin{figure}[htb]
\epsfig{figure=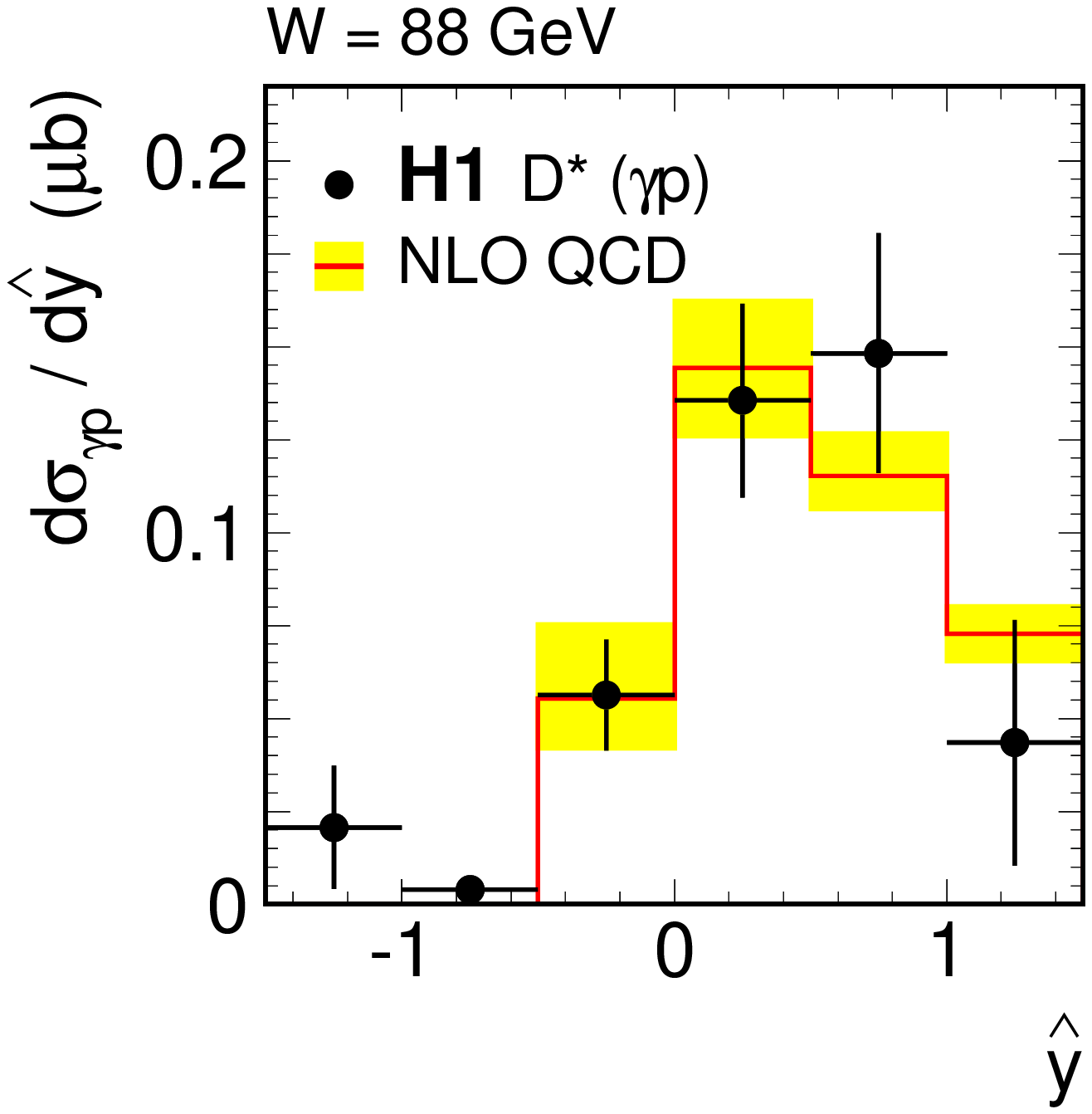 ,width=0.32\linewidth}
\hfill
\epsfig{figure=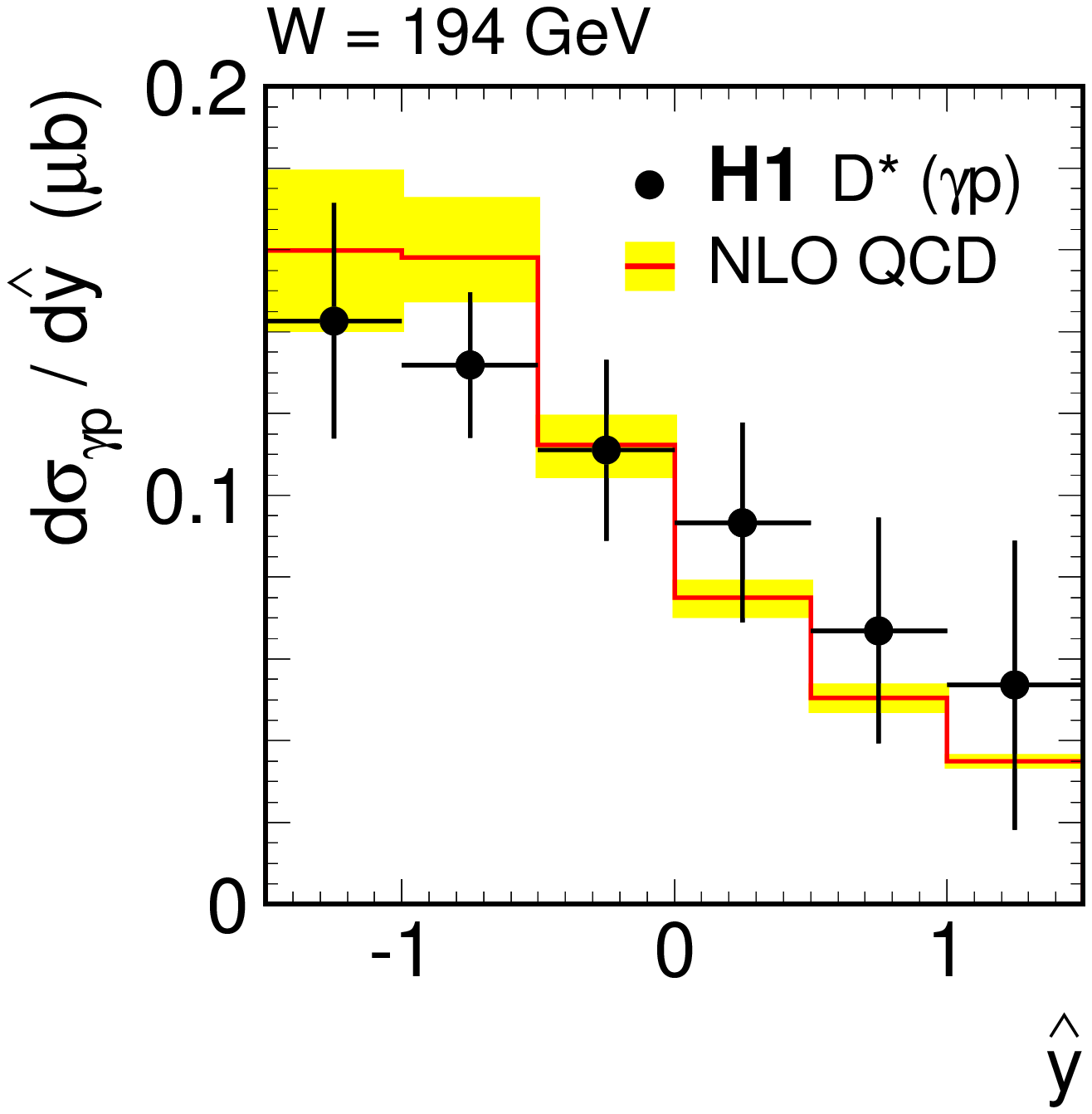 ,width=0.32\linewidth}
\hfill
\epsfig{figure=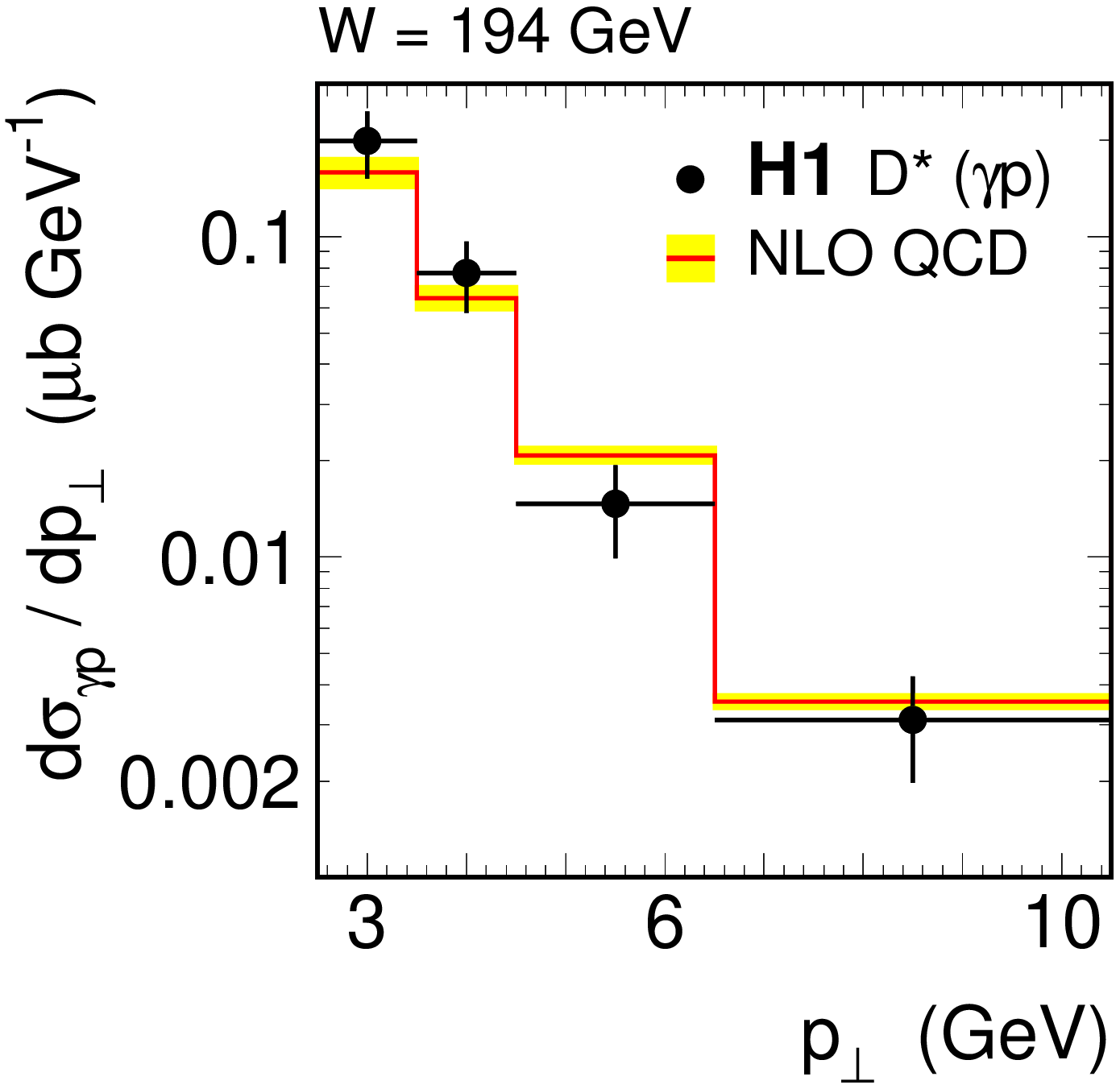 ,width=0.32\linewidth}

\begin{picture}(1,1)(0,0)
\put(100,140){\begin{footnotesize}(a)\end{footnotesize}}
\put(250,140){\begin{footnotesize}(b)\end{footnotesize}}
\put(400,140){\begin{footnotesize}(c)\end{footnotesize}}
\end{picture}

\vspace{-1cm}
\caption{Tagged PHP \ds\ cross section differential in rapidity (a)  (b) and \ptr\ (c).
}
\label{f:h196}
\end{figure}

The H1 Collaboration has measured open charm PHP selected using  the 44 ({\it 33}) m taggers~\cite{h196}. The kinematic region of the measurement is:
$Q^2<0.009 {\it (0.01)} $GeV$^2$, $<W>=88 {\it (194)}~$GeV, 
$p_T(\ds)>2 {\it (2.5)}~$GeV and $\mid {\rm rapidity}(\ds) \mid<$ 1.5. The data are compared  
in Figure~\ref{f:h196} with the FMNR prediction using  pdf's MRST1 (p, $\alpha_s$=0.1175), 
 GRV-HO ($\gamma$), $\epsilon=0.035$, 
 $\mu_R=\sqrt{m_c^2+p_T^2}~ ;~ m_c=1.3-1.7~$GeV$~ ;~ \mu_F=2\mu_R$. 
 This calculations used an old value of $f(c\rightarrow \ds)$, 15\% higher  than the latest
result (0.235). Also the use of a larger value of $\alpha_s$ than in the case of the ZEUS analysis 
produces higher values of the prediction.
 The H1 Collaboration finds good agreement with FMNR and uses this calculation 
 to extract $x_g g(x_g)$.

\vspace{-0.6cm}


\section{Open charm electroproduction}

\vspace{-0.2cm}

  Earlier HERA ($\sqrt{s}=300$~GeV) and  fixed target data ($\sqrt{s}\sim 30$~GeV) 
show that open charm electroproduction is dominated by PGF, thus providing  
 direct sensitivity to the gluon pdf in the $p$. FO NLO pQCD calculations are available 
 in the form of a MC integrator (HVQDIS)~\cite{HVQDIS}. 
Only light quarks and the gluon are present in the 
$p$ as pdf's.  Since open charm electroproduction is sensitive to 
the gluon, this process allows  a test  of factorization and in particular 
of the universality of the gluon pdf, comparing the data with NLO predictions
which use as an input  the pdf's obtained from fits to inclusive \ftwo. 
The same fragmentation limitations mentioned before for PHP apply here.

%
\begin{figure}[htbp]
\begin{center}
\epsfig{figure=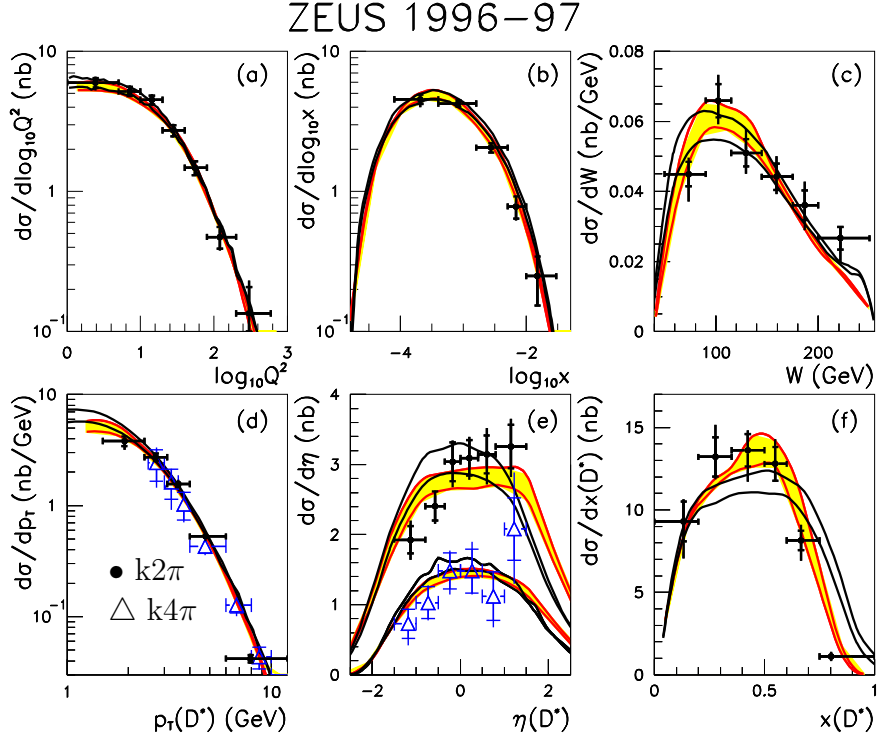 ,width=0.8\linewidth,clip=}
\end{center}
\begin{picture}(1,1)(0,0)
\put(100,60){$\bullet$ k2$\pi$} 
\put(100,45){$\triangle$  k4$\pi$}
\end{picture}
\caption{
Differential cross sections for \ds\ production from the $K2\pi$ final state (solid dots) and from the $K4\pi$ channel (open triangles).
The open  band corresponds to the
standard Peterson fragmentation function. The  shaded band shows the NLO reweighted JETSET MC. \xd\ (f) is defined as $2p^*(\ds)/W$, where $p^*$ is in the $\gamma^*$-$p$ CMS frame.
}\label{f:ds_diff_97}
\end{figure}

\begin{figure}[htb]
\begin{minipage}{0.7\linewidth}
\epsfig{file=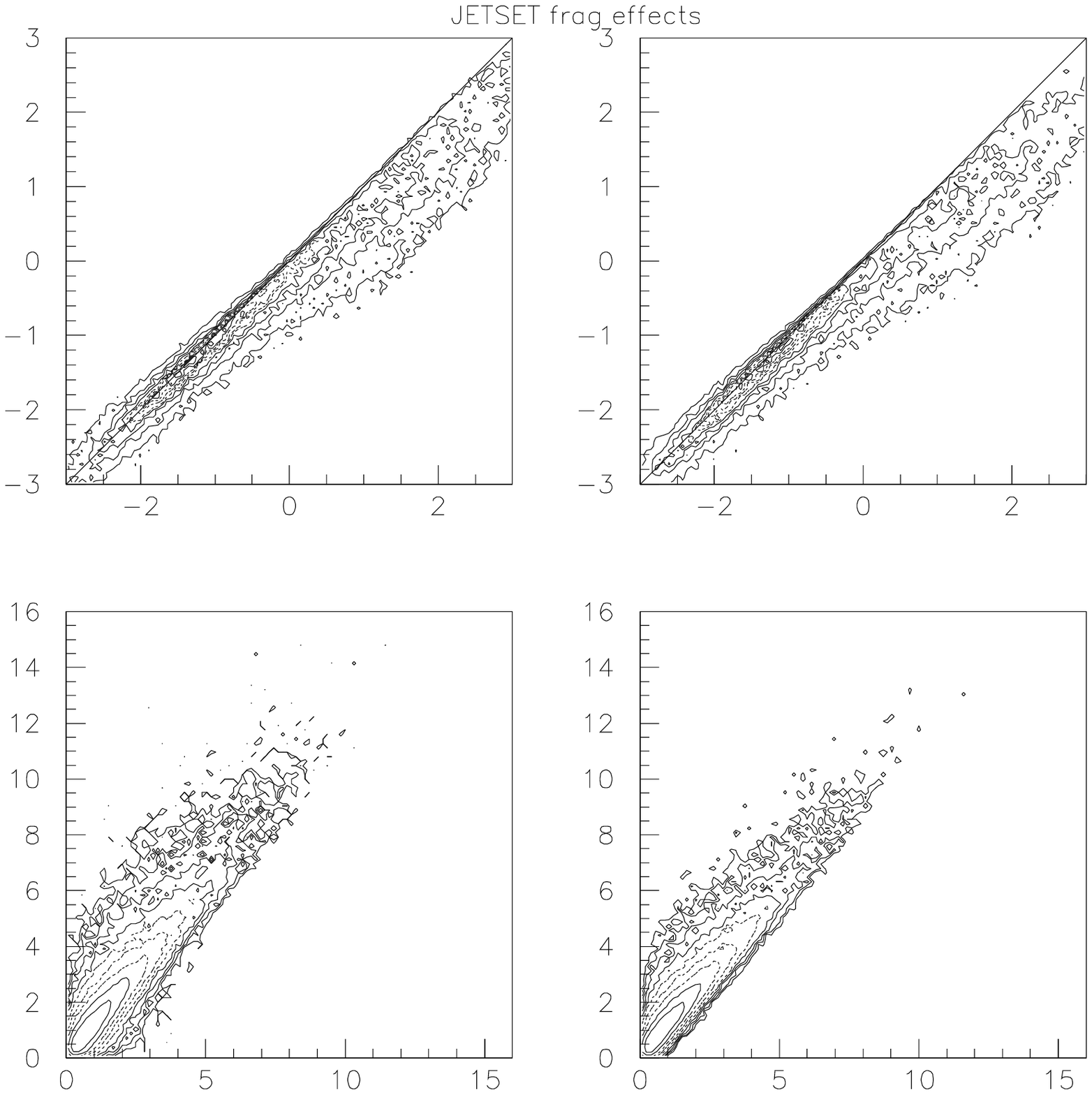,width=1\linewidth,clip=0}
\begin{picture}(1,1)(0,0)
\put(0,-180){\White{\rule{2000pt}{200pt}}}
\end{picture}
\begin{picture}(1,1)(-50,0)
\put(20,10){$\eta(\ds)$}
\put(180,10){$\eta(\ds)$}

\put(-60,90){$\eta(c)$}
\put(95,90){$\eta(c)$}

\put(-12,120){(a) before PS}
\put(140,120){(b) after PS}

\put(-28,33){\rule{2pt}{122pt}}
\put(-28,33){\rule{122pt}{2pt}}
\put(130,33){\rule{2pt}{122pt}}
\put(130,33){\rule{122pt}{2pt}}

\put(260,120){(c)}
\end{picture}
\caption{
$\eta(c)$-$\eta(\ds)$ correlations from JETSET before (a) and after (b) the parton shower.
Sketch (c) to illustrate the similarity of PGF and  $e^+e^- \rightarrow q\bar{q} g$.
}
\label{f:etacorr}
\end{minipage}
\begin{minipage}{0.29\linewidth}
\epsfig{file=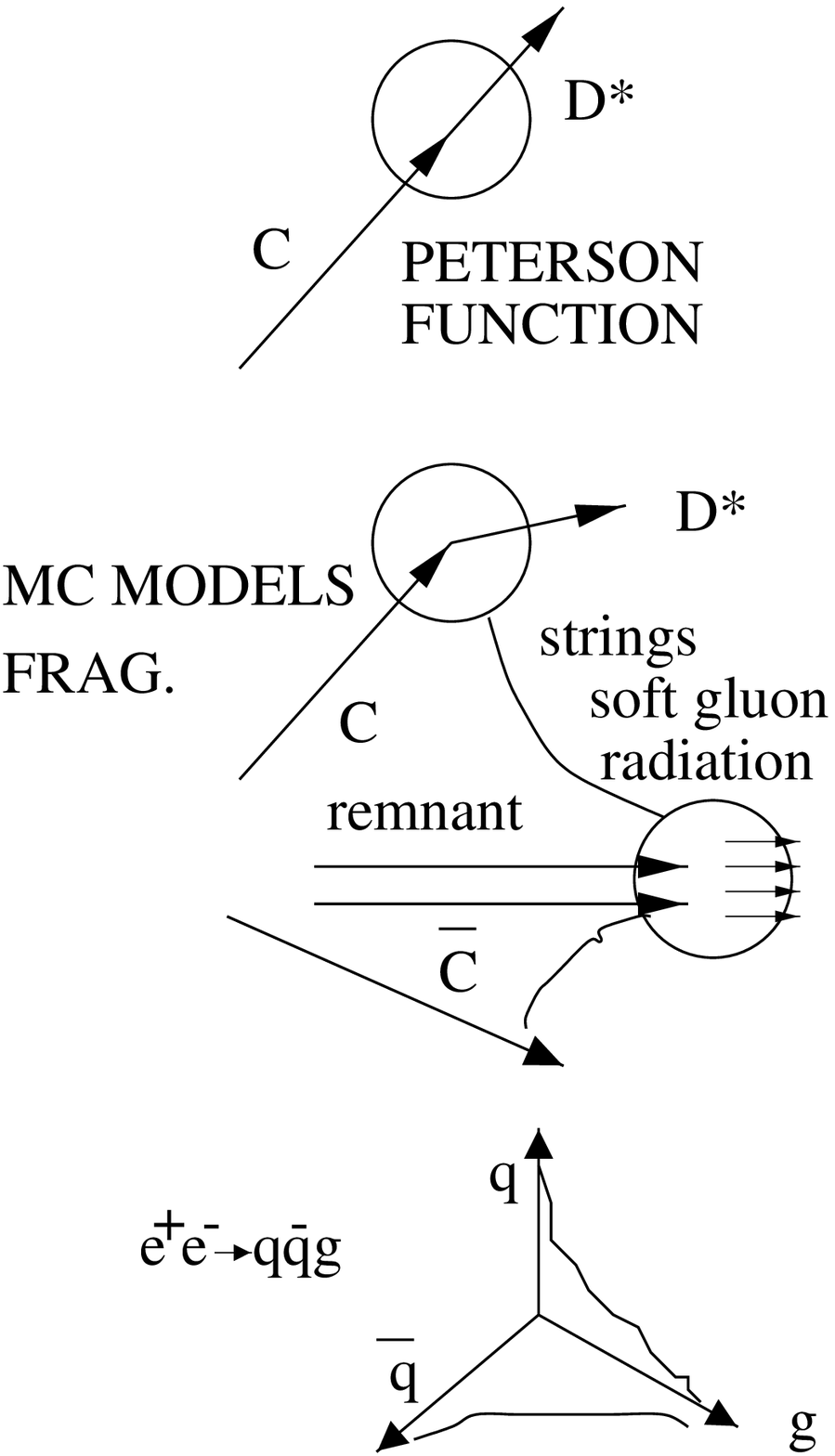 ,width=1\linewidth,clip=}
\end{minipage}
\end{figure}

ZEUS has measured \ds\ production in DIS using  
${\cal L}\sim 37~ $pb$^{-1}$ at $\sqrt{s}=300$~GeV in the kinematic region 
$1<Q^2<600~$GeV$^2$, $0.02<y<0.7$,  $1.5({\it 2.5})<p_T(\ds)<15~$GeV and  $\mid \eta(\ds) \mid<$1.5 \cite{zds9697}.
 The differential cross sections are shown in Figure~\ref{f:ds_diff_97}.
The two decays used, $\ds \rightarrow K2\pi {\it (K4\pi)}$, are in agreement as can be seen in the
\ptds\ distribution.
The settings of  the NLO HVQDIS prediction are: pdf ZEUS NLO fit ($\alpha_s$=0.119), 
$\epsilon = 0.035$ (in the lab), $\mu_R=\mu_F=\sqrt{4m_c^2+Q^2}$, $m_c=1.3-1.5~$GeV.
 An old  value of $f(c\rightarrow\dsp)$, 9 \% lower than the latest result 
(0.235), was used.  The comparison of the data with this prediction (open band) yields the following observations: 
\begin{itemize}
\item Good agreement with HVQDIS in $Q^2$, $x_{Bj}$ and $W$.
\item HVQDIS is shifted with respect to  the data in the \etads, \xd\ distributions. 
\end{itemize}
However, the shaded  band, obtained by folding JETSET and the same  NLO calculation, describes 
better the data. The NLO calculation is folded with JETSET by means of reweighting
the $pt(c)$,$\eta (c)$ two dimensional  distribution of a LO MC after the hard interaction and letting JETSET  hadronize. 
 A shift into the forward direction due to the string model is observed.
 A similar picture is obtained using  HERWIG or ARIADNE.
 Figure ~\ref{f:etacorr} shows that the shift in $\eta$ coming from the full fragmentation model (PS+string model) is  similar to  the effect coming from the string model alone.
 This shift is expected in models which include the interaction of the colour charges 
in the final state since the colour configuration is similar to $e^+e^- \rightarrow q\bar{q} g$, 
where the string effect was observed. In PGF processes  the remnant is acting as a   colour octet which drags the
$c$, $\bar{c}$ into the forward region.

ZEUS has measured cross sections of $e^-$ coming from semileptonic decays of charm  
in DIS using ${\cal L }\sim 34$ pb$^{-1}$~\cite{semileptonic}. This measurement relies on the 
 dE/dx  deposition of the tracks in the CTD to separate the $e^-$ signal from 
the hadronic background. The kinematic region is:
$1.2<p_e<5~$GeV, $\mid \eta_e \mid <1.1$, $1<Q^2< 1000~ $GeV$^2$, $0.03<y<0.7$.
The method is complementary to the  \ds\ analysis since, on one hand, 
 it profits from the larger branching fraction to go to higher $Q^2$ and, on the other hand,
it probes harder $c$ quarks, providing a $p_T$ enriched sample. The price to
pay is that it covers a smaller region of the $c$ quark phase space.
The data are in  agreement with NLO HVQDIS, which in this case uses the settings:
 pdf GRV94 H0 ($\alpha_s=0.111$), $\epsilon = 0.035 $,  $\mu_R=\mu_F=\sqrt{4m_c^2+Q^2}$ and  $m_c=1.3-1.7~$GeV.


\begin{figure}[htb]
\begin{center}
\epsfig{figure=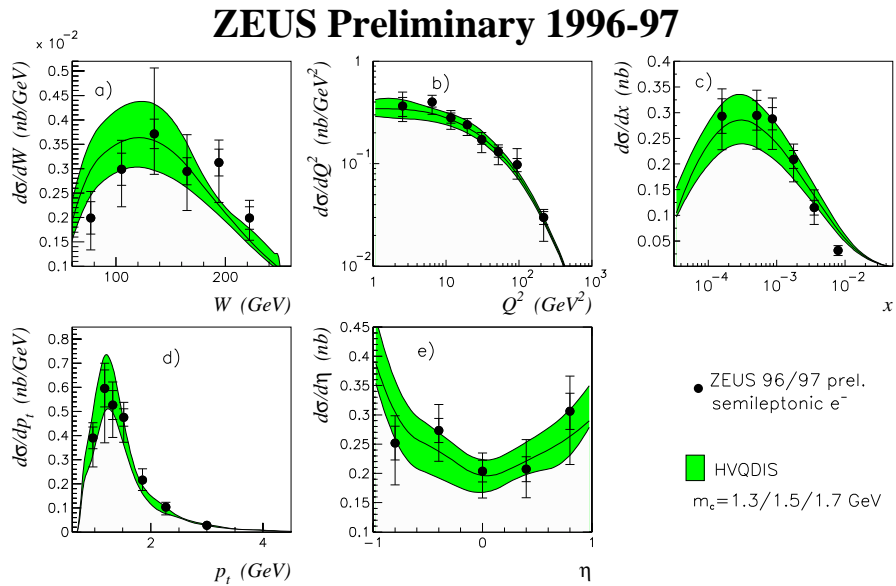 ,width=0.8\linewidth,clip=}
\end{center}


\caption{Semileptonic $e^-$ differential cross sections.
}\label{f:semilep}
\end{figure}

\begin{figure}[htb]
\begin{minipage}{0.32\linewidth}
\begin{center}
\epsfig{figure=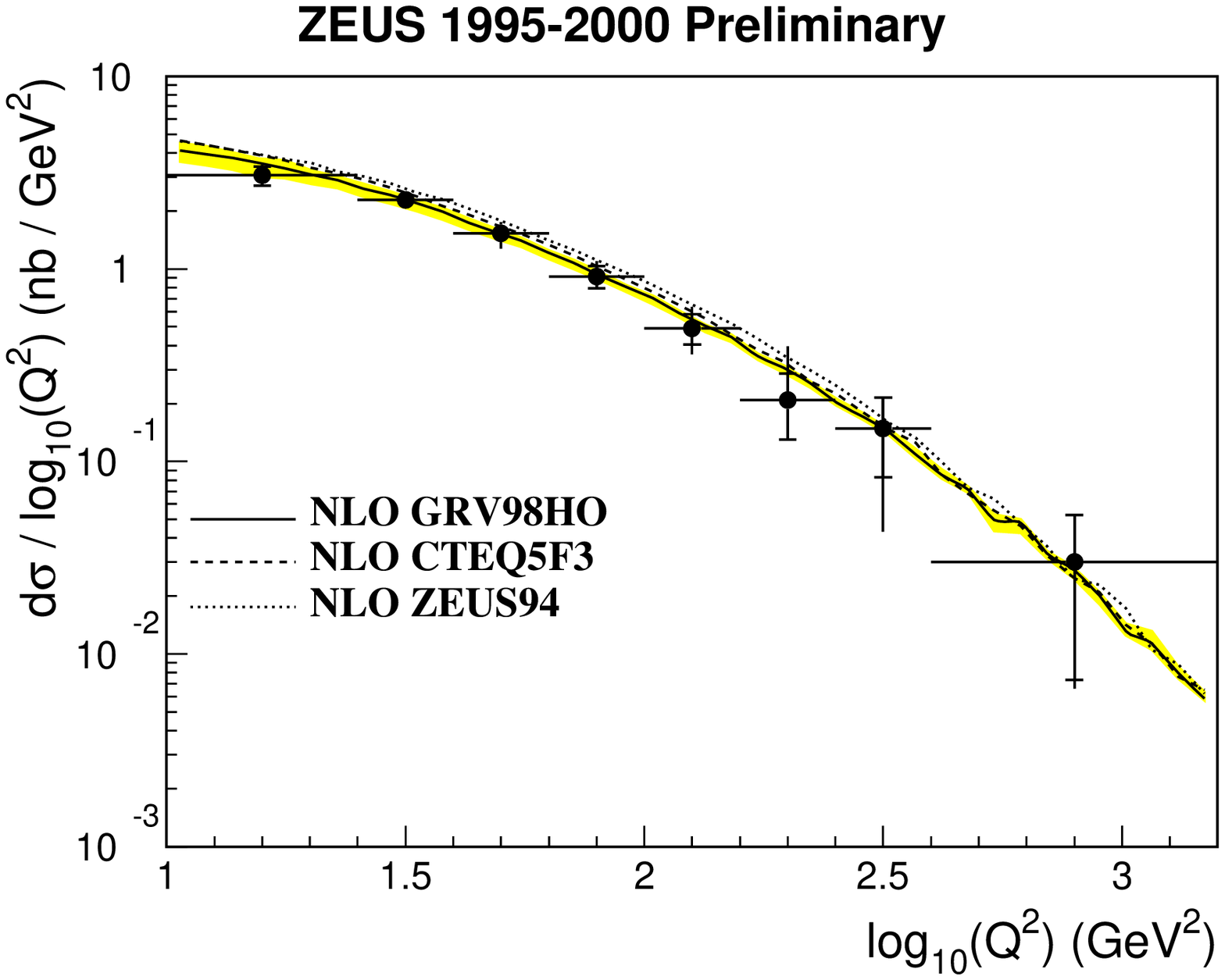 ,width=1.\linewidth}
\end{center}
\end{minipage}
\hfill
\begin{minipage}{0.32\linewidth}
\begin{center}
\epsfig{figure=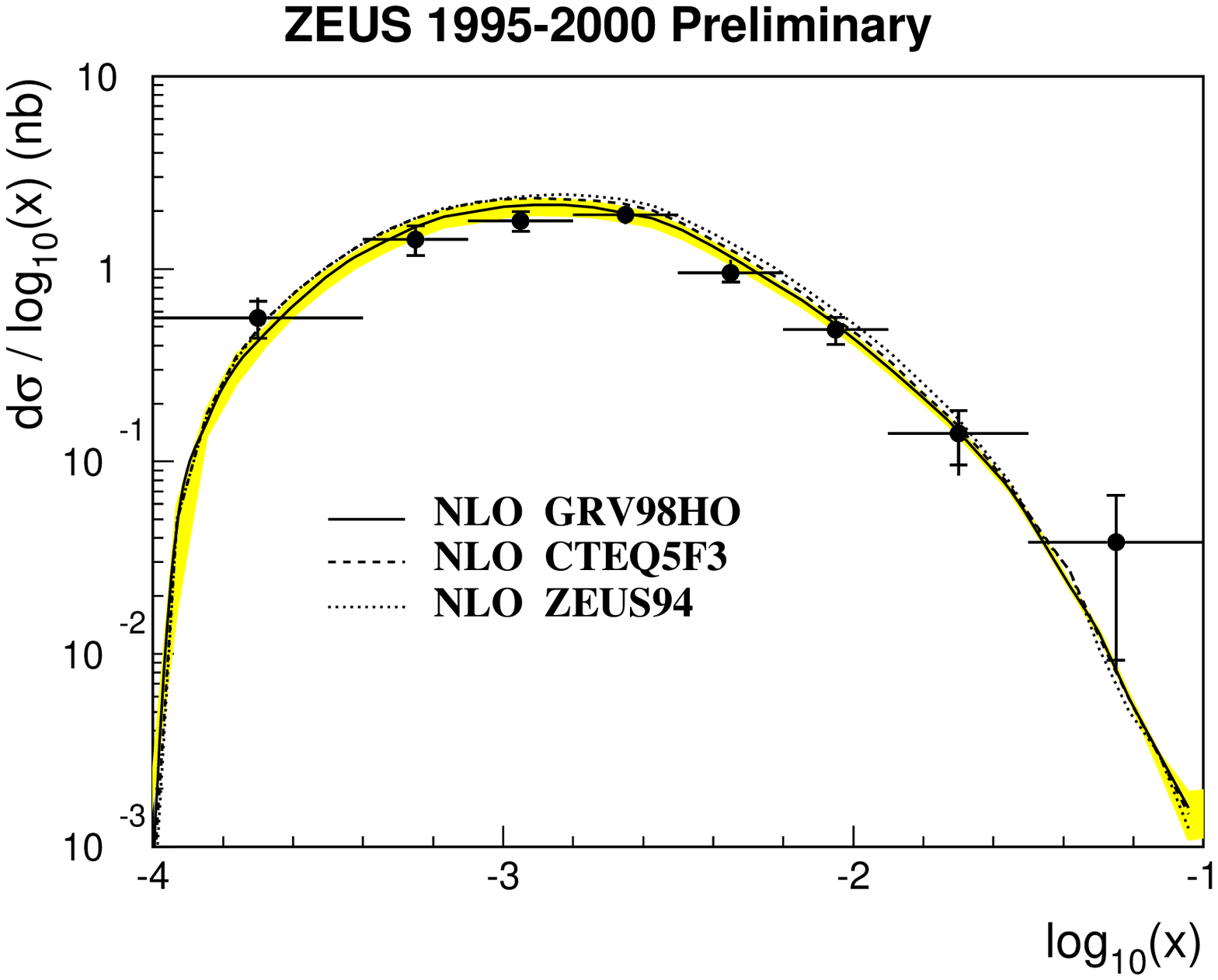 ,width=1.\linewidth}
\end{center}
\end{minipage}
\begin{minipage}{0.32\linewidth}
\begin{center}
\begin{footnotesize}
\begin{tabular}{c c l}
pdf      & $\Lambda^4_{QCD}$ & $\sigma$(nb) \\
         &  (MeV)            &  $m_c$=1.3-1.6\\\hline
GRV98    & 257 & 2.71-2.27 \\
CTEQF53  & 326 &  2.95-2.48 \\
ZEUS NLO & 404 &  3.2-2.63\\
\end{tabular}
\end{footnotesize}
\end{center}
\end{minipage}

\begin{picture}(1,1)(0,0)
\put(100,90){(a)}
\put(260,90){(b)}
\end{picture}
\vspace{-0.3cm}
\caption{
\ds\  cross sections differential in $log_{10}(Q^2)$ (a) and  
 $log_{10}(x_{Bj})$ (b).
}\label{f:ds318} 
\end{figure}

\begin{figure}[htb]
\begin{minipage}{0.49\linewidth}
\caption{ FO NLO pQCD (labeled Exact) is compared with two VFNS predictions and with 
H1 and ZEUS data. Figure taken from ~\cite{chuvakin}.
}\label{f:harris}
\end{minipage}
\hfill
\begin{minipage}{0.4\linewidth}
\epsfig{figure=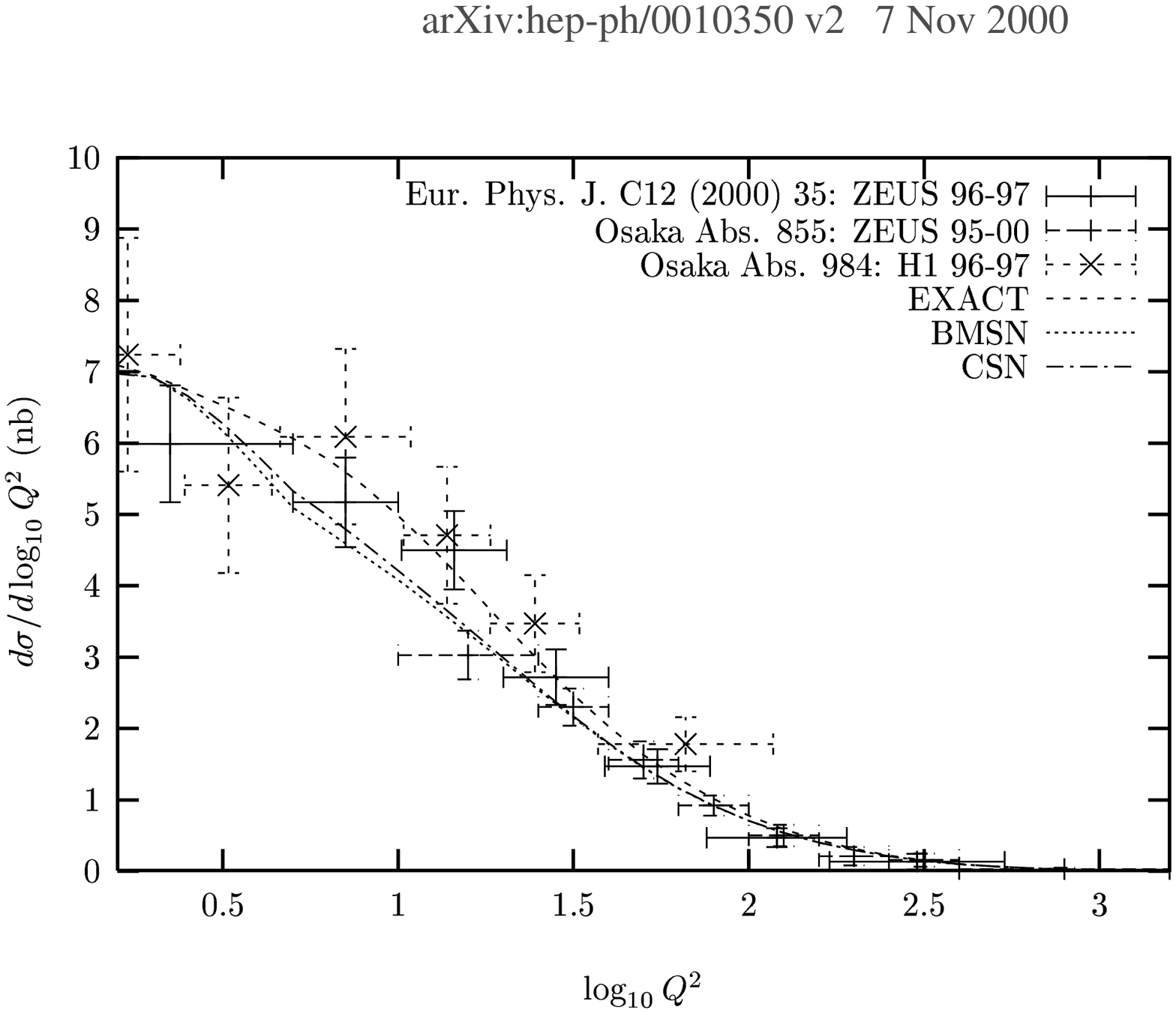,angle=270,width=1.1\linewidth,clip=}
\begin{picture}(0,0)(1,1)
\put(0,150){\White{\rule{200pt}{20pt}}}
\end{picture}
\end{minipage}
\end{figure}

ZEUS has  also measured \ds\ production cross sections integrating two samples at 
different $\sqrt{s}$,
${\cal L}_{\sqrt{s}=300,318} \sim 45, 38~ $pb$^{-1}$~\cite{ds318}.
 The integrated cross section 
in the kinematic region  $10<Q^2< 1000 ~$GeV$^2$, $0.04<y<0.95$, $1.5<p_T(\ds)<15~$GeV 
and $\mid \eta(\ds) \mid <1.5$ is
$\sigma(e^+p \rightarrow e^+D^{*\pm}X) = 
2.33 \pm 0.12(\mbox{stat.})^{+0.11}_{-0.07}(\mbox{syst.}) \mbox{ nb}$.
 This result and the differential cross sections compare reasonably
 well with NLO HVQDIS using 
 three different pdf's, see Figure~\ref{f:ds318}. 
 Although the measurement limits have
 been pushed towards  high $Q^2$ and  $x_{Bj}$, MC studies show that 
there is no  sensitivity to intrinsic charm (IC) yet. Another issue in this region is 
 whether variable flavour number schemes (VFNS), which include a c pdf arising from the 
resummation of large $log(Q^2/m_c)$ can give a better description of the data than that of 
the FO NLO pQCD predictions. Little difference between FFNS and VFNS's was found in 
~\cite{chuvakin} until $x_{Bj}>0.1$ and high $Q^2$.
 However, a slight difference appears at $Q^2\sim 10~$GeV$^2$, Figure~\ref{f:harris}.
 Even if the effect is not an artifact
introduced by the use of HVQDIS to produce  VFNS predictions differential in the c phase space 
 or by the normalization of all  curves at $Q^2=m_c^2$, the question is whether it can be 
 distinguished from  fragmentation and/or mass uncertainties which 
 are of the same order.

\vspace{-0.8cm}

\vspace{-0.8cm}

The H1 Collaboration has performed a measurement of \ds\ production 
in DIS using  ${\cal L}\sim 18~ $pb$^{-1}$~\cite{h197}. The integrated cross section 
in the kinematic region $1<Q^2<100~$GeV$^2$, $0.05<y<0.7$, $1.5~GeV<p_T(\ds)$ 
and $\mid \eta(\ds) \mid <1.5$ is:
$\sigma_{KIN}(e^+p \rightarrow e^+D^{*\pm}X) = 
8.50 \pm 0.42(\mbox{stat.})^{+1.02}_{-0.76}(\mbox{syst.}) \pm 0.65(\mbox{mod.}) \mbox{ nb}$.
 This can be directly compared with the ZEUS measurement~\cite{zds9697} in the similar kinematic region
 $1<Q^2<{\it 600}~$GeV$^2$, ${\it 0.02}<y<0.7$, $1.5 <p_T(\ds)<15~$GeV, $\mid \eta(\ds) \mid <1.5$: 
$\sigma_{KIN}(e^+p \rightarrow e^+D^{*\pm}X) = 
8.31 \pm 0.31(\mbox{stat.})^{+0.3}_{-0.5}(\mbox{syst.})  \mbox{ nb}$ using HVQDIS to interpolate
between the two measurements. 
\begin{equation}
R_{KIN}=\frac{\sigma_{ZEUS KIN}^{HVQDIS}}{\sigma_{H1 KIN}^{HVQDIS}}=1.11~~\Rightarrow~~\frac{R_{KIN}^{DATA}}{R_{KIN}^{HVQDIS}}= 0.88^{+0.12}_{-0.15}, 
\end{equation}
 which is  compatible with 1.  
This, and the comparison of  the differential cross sections in $log_{10}(Q^2)$, see Figure~\ref{f:harris}, support the fact that  H1 and ZEUS data agree within errors.  

\begin{figure}[htb]
\begin{minipage}{0.49\linewidth}
\begin{center}
\epsfig{figure=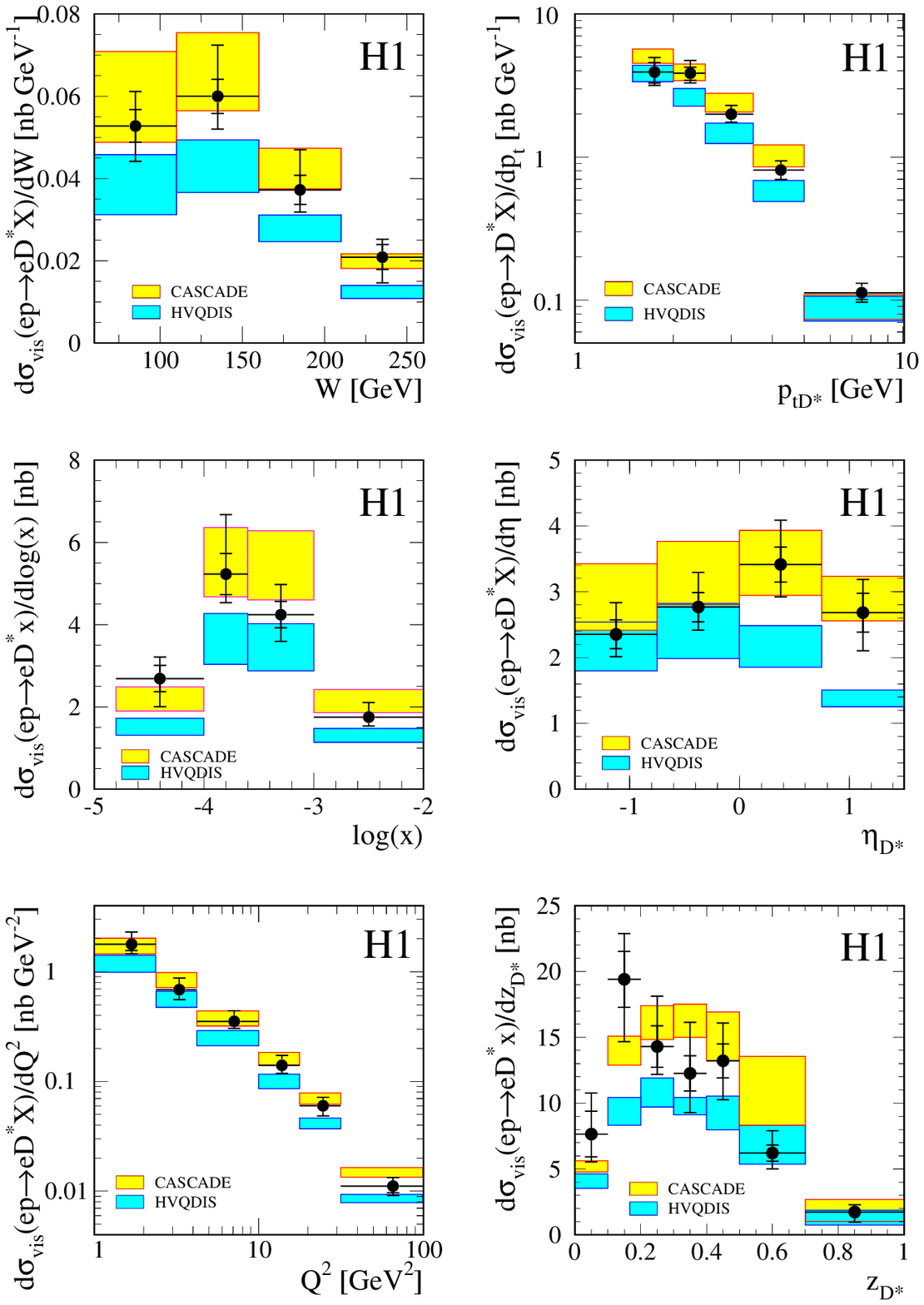 ,width=0.95\linewidth}
\end{center}
\begin{footnotesize}
\begin{picture}(1,1)(0,0)
\put(88,285){(a)}
\put(193,285){(d)}
\put(88,190){(b)}
\put(193,190){(e)}
\put(88,90){(c)}
\put(193,90){(f)}

\put(410,220){(g)}
\end{picture}
\end{footnotesize}
\end{minipage}
\hfill
\begin{minipage}{0.49\linewidth}
\centerline{\epsfig{figure=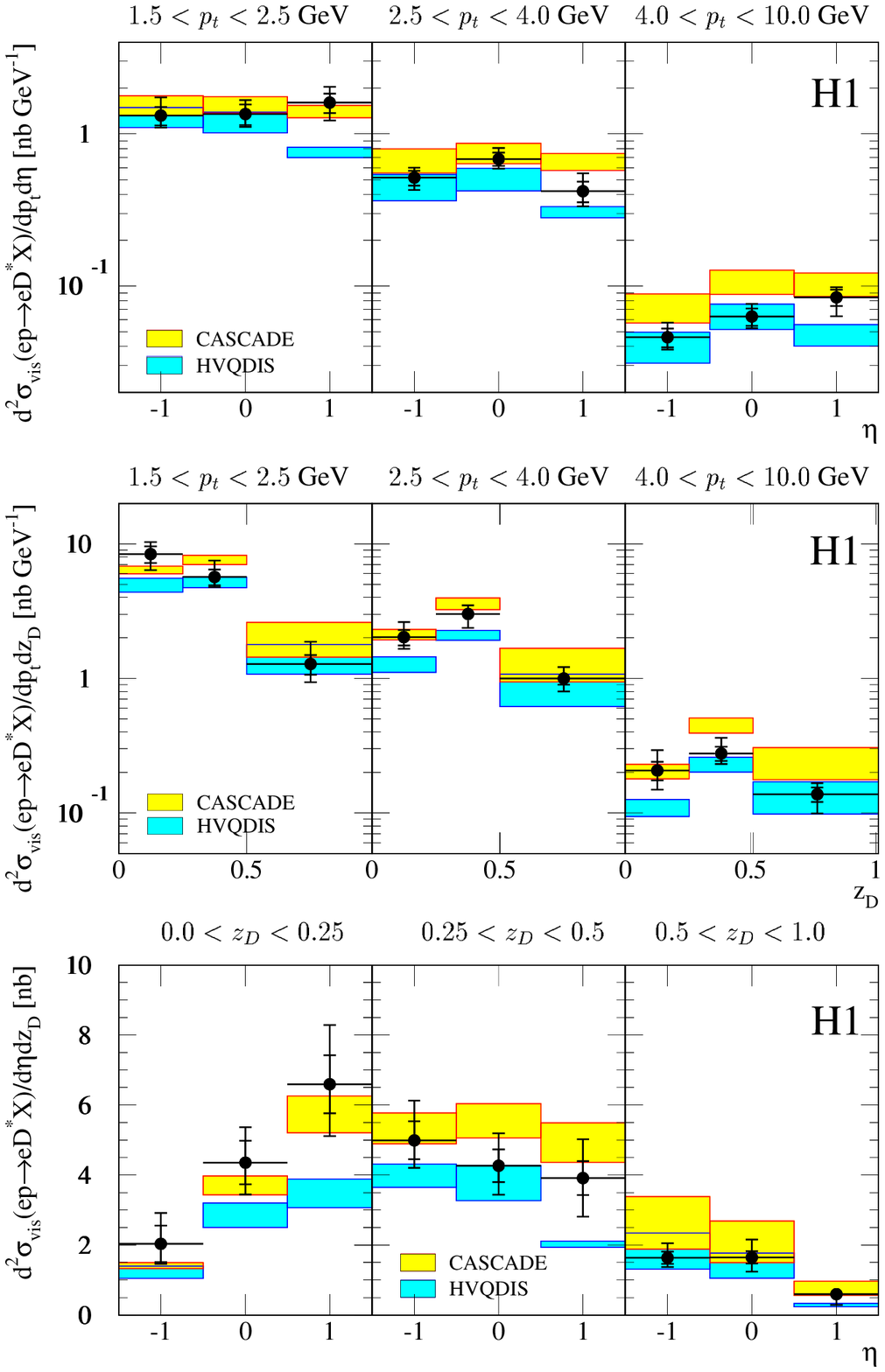 ,bbllx=90,bblly=500,bburx=500,bbury=750,width=0.9\linewidth,clip=}}
\begin{picture}(1,1)(0,0)
\put(0,-472){\White{\rule{1000pt}{500pt}}}
\end{picture}
\vspace{-1cm}
\caption{
Differential \ds\ cross sections vs $W$ (a), $x_{Bj}$ (b), $Q^2$ (c), \ptds\ (d), \etads\ (e), 
$z_{D^\ast}=(E{\rm -}P_z)^{D^\ast}/(E{\rm -}P_z)^{\gamma^*}$ (f) compared with HVQDIS and CASCADE predictions (see text). Double differential cross section in 
\ptds\ and \etads\ (g) compared with the same calculations. 
}\label{f:h197}
\end{minipage}
\end{figure}

 H1's data are  compared with HVQDIS using pdf GRV98 ($\alpha_s$=0.114), $m_c$=1.3-1.5~GeV. 
In addition to the standard Peterson (in  $\gamma^*p$ CMS) with  $\epsilon$=0.035-0.1,
a transverse momentum with respect to the charm quark is given to the \ds\ meson, 
 according to the function $exp(-{\alpha p_t^2})$ with $<p_t^2> \approx$350~MeV.  This results in a prediction of $7.02-5.17$ nb. 
 The H1 analysis also compares the data with CASCADE~\cite{CASCADE}, which implements the CCFM evolution. 
 In this model an unintegrated gluon density, fitted to H1's \ftwo, and $m_c$=1.3-1.5~GeV is used.
 The fragmentation is performed by JETSET with $\epsilon$=0.035-0.1 yielding $10.77-8.04$ nb.
 The measured single and double differential cross sections are compared to the same calculations in  Figure~\ref{f:h197}. CASCADE shows better agreement in normalization, while the previous  HVQDIS prediction is too low 
at forward $\eta$, low $z_{D^\ast}$ (Figures~\ref{f:h197} (e) and (f)).
The double differential cross section (Figure~\ref{f:h197} (g)) shows that
 the disagreement with HVQDIS is concentrated at low \ptds\ and  forward \etads, which is correlated
 with low $z_{D^\ast}$. Moreover there is a poor agreement with CASCADE at high \ptds.
\vspace{-0.5cm}

\vspace{-1.5cm}

\subsection{Extraction of \ftwoccb}
\begin{figure}[htb]
\begin{minipage}{0.49\linewidth}
 \epsfig{figure=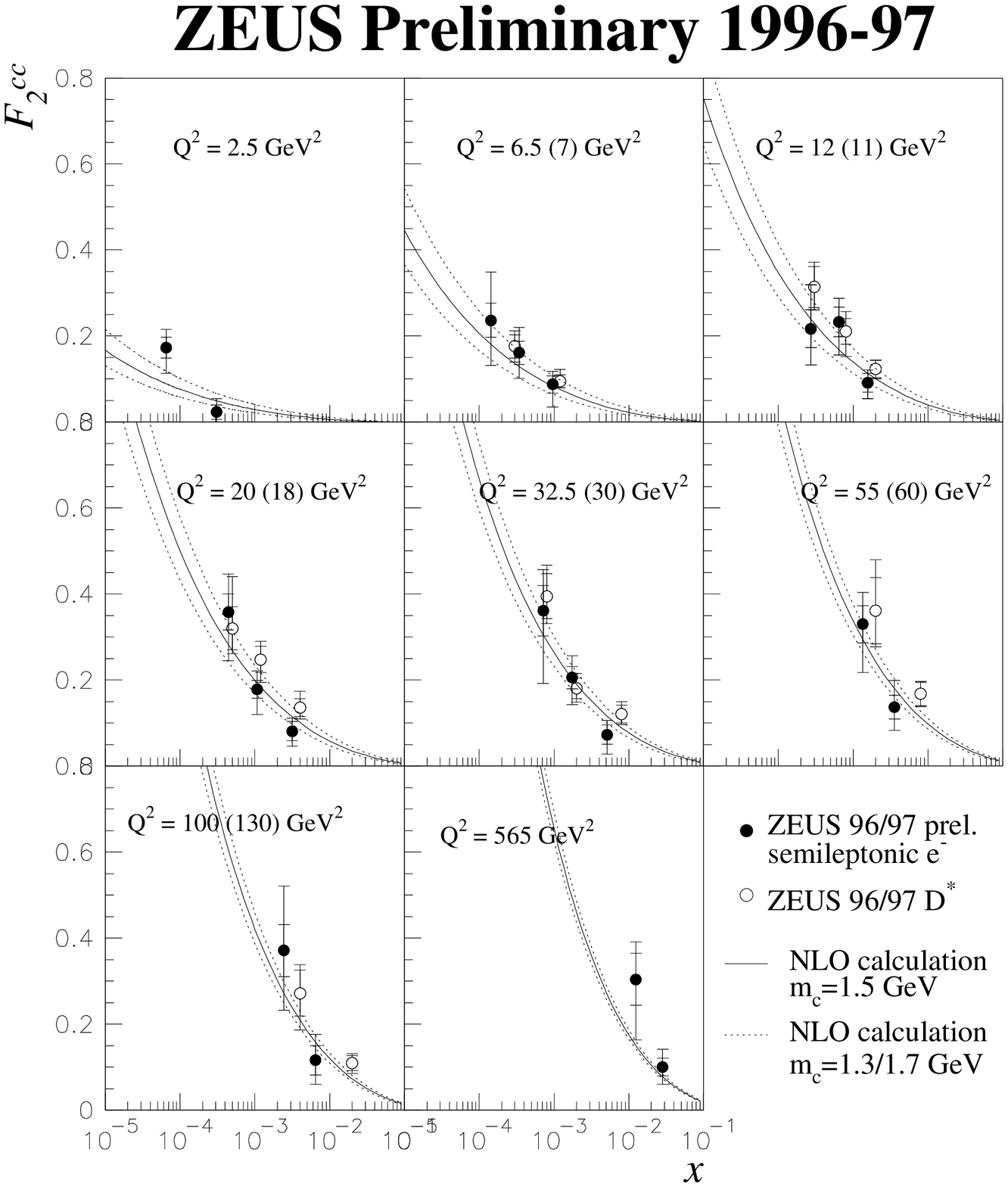 ,width=0.9\linewidth}
\end{minipage}
\begin{minipage}{0.49\linewidth}
\begin{center}
 \epsfig{figure=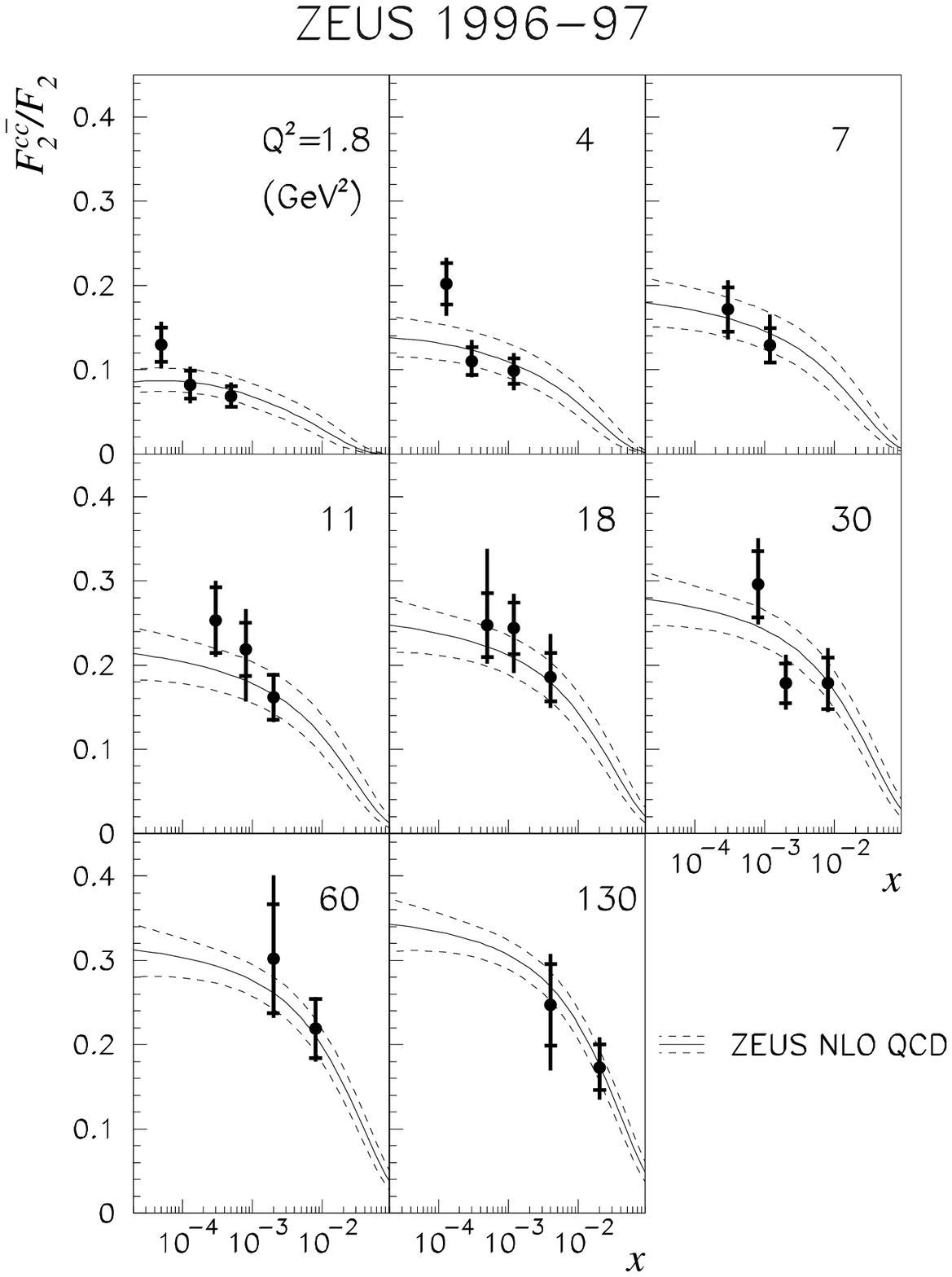 ,width=0.8\linewidth}
\end{center}
\end{minipage}
\begin{minipage}{0.49\linewidth}
 \epsfig{figure=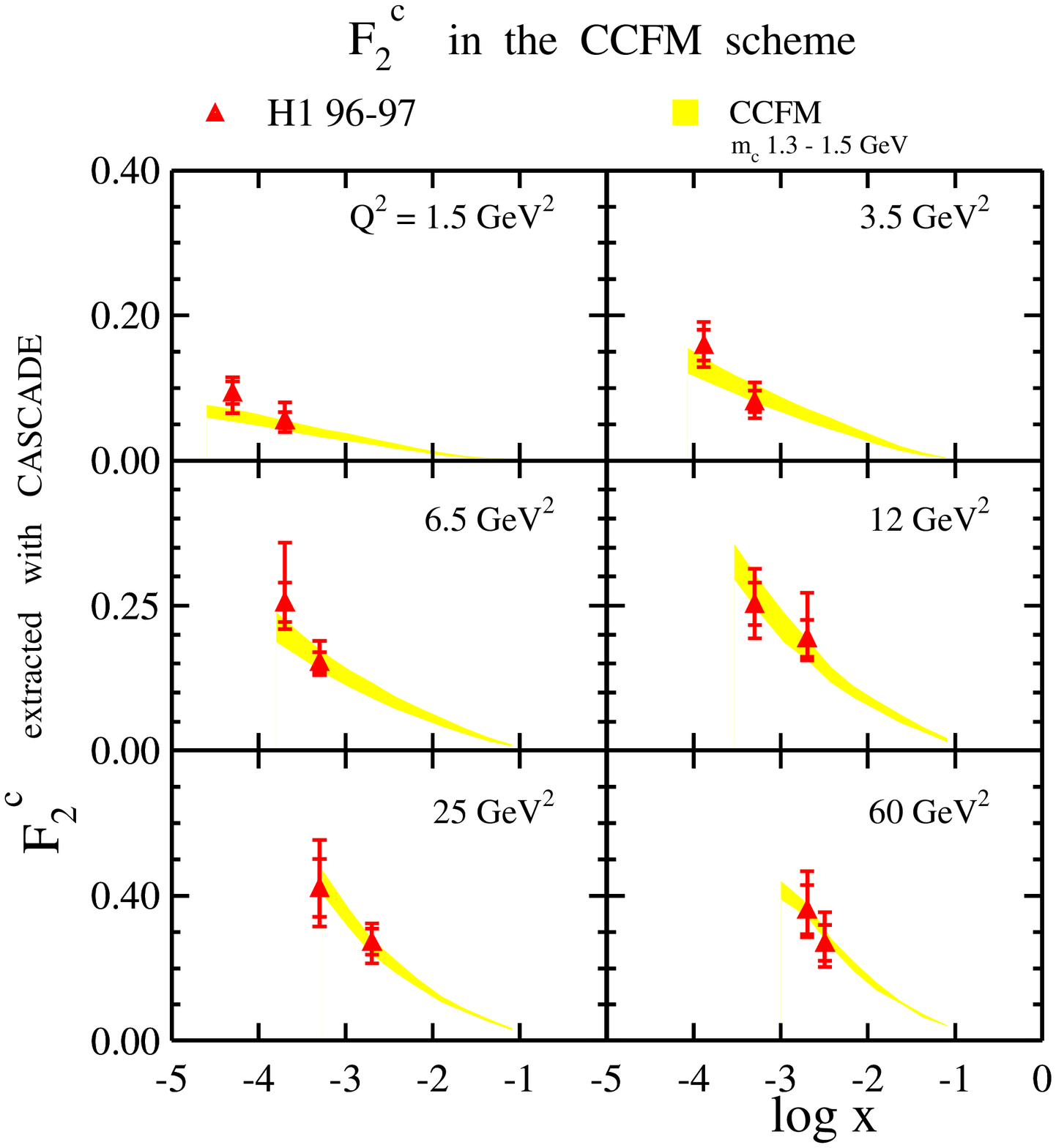 ,width=0.8\linewidth}
\end{minipage}
\begin{minipage}{0.49\linewidth}
 \epsfig{figure=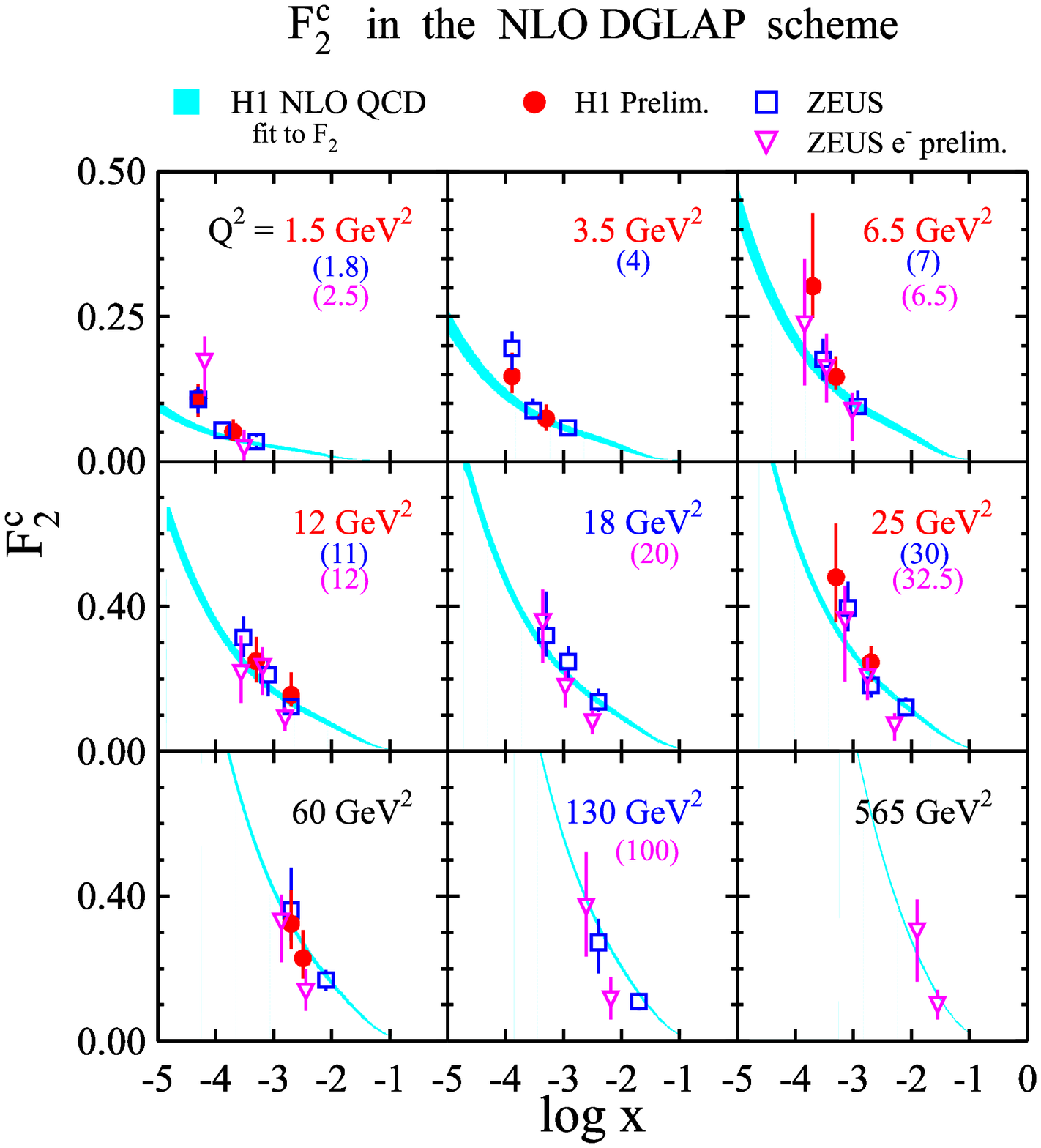 ,width=0.85\linewidth}
\end{minipage}

\begin{Large}
\begin{picture}(1,1)(0,0)
\put(-20,400){(a)}
\put(430,400){(b)}
\put(-20,150){(c)}
\put(430,150){(d)}
\put(0,-280){\White{\rule{100pt}{140pt}}}
\end{picture}
\end{Large}

\vspace{-0.5cm}

\caption{
Measured \ftwoccb\ from semileptonic data~\cite{semileptonic} vs. $x_{Bj}$ at fixed $Q^2$ values (a) compared
to the model used for the extrapolation (HVQDIS, GRV94). Ratio   of the measured \ftwoccb\ 
from \ds\ data~\cite{zds9697} extrapolated with  (HVQDIS, ZEUSNLO) over \ftwo\ vs. $x_{Bj}$ at fixed $Q^2$ values (b). 
Measured \ftwoccb\ from
 \ds\ data~\cite{h197} vs. $x_{Bj}$ at fixed $Q^2$ values   compared
to the model used for the extrapolation (CASCADE) (c).
The data of (a) and (b) 
compared with the H1 data~\cite{h197} extracted with HVQDIS,GRV98 (d). 
}\label{f:f2c}
\end{figure}

 \ftwoccb\ can be defined in terms of the double differential cross section integrating in the 
full c quark phase space:
$\frac{d^2\sigma^{c\bar{c}} (x, Q^2)}{dxdQ^2} = 
\frac{2\pi\alpha^2}{x Q^4}
\{ [1+(1-y)^2] F_2^{c\bar{c}}(x, Q^2) - y^2 F_L^{c\bar{c}} \}$. 
Several assumptions are made in the extraction:
\begin{itemize}
\item The contribution of   $F_L^{c\bar{c}}$ ($<1\%$, according to the FO NLO calculations) 
is neglected.
\item Bound charm ($<$2.5-4.5 $\%$~\cite{zds9697}) is neglected.
\item  $f(c\rightarrow \dsp)$ from $e^{+}e^{-}$ is valid at HERA.
\item The extrapolation factors, $\sigma_{INT} \over{\sigma_{KIN}}$, are well 
 described by the model used to extrapolate.
\end{itemize}
 The relevance of the last assumption is illustrated by the size of the extrapolation factors,
which go from $\leq$2 at high $Q^2$ to $\leq$4 at low $Q^2$ for \ds\ data and can be as large as 
  $\leq$20 at low $Q^2$ for $e^-$ data.
The actually applied procedure to extrapolate outside the kinematic region (KIN) in $\ptr,\eta$ is:  
\begin{equation}
\label{eq:extra}
\ftwoccb_{meas}(\qsq,x_{Bj})= \ftwoccb_{theo}(\qsq,x_{Bj}) \times{{\sigma_{KIN}^{meas}(\qsq,y)}\over{\sigma_{KIN}^{theo}(\qsq,y)}}  
\end{equation}
The extrapolation is not strongly affected by  the fragmentation models because the procedure
is not very sensitive to the shape of the $\eta$ distribution.
 The information contained in  the  \ftwoccb\ plots in Figure~\ref{f:f2c} can be classified according 
 to its dependence on the extrapolation:
\begin{enumerate}
\item 
 Since $\ftwoccb_{meas}/\ftwoccb_{theo}=\sigma_{KIN}^{meas}/\sigma_{KIN}^{theo}$  by construction (see eq.~\ref{eq:extra}), the   comparison of  $\ftwoccb_{meas}$ with the $\ftwoccb_{theo}$   that was  used to extrapolate represents the level of agreement at the cross section level and is independent of the extrapolation.
   The semileptonic double differential  cross section $\sigma^{meas}_{KIN}(Q^2,x_{Bj})$ is
  well described by HVQDIS (GRV94) as can be seen in Figure ~\ref{f:f2c} (a).
   The same is true for the \ds\ data and HVQDIS (ZEUSNLO), Figure ~\ref{f:f2c} (b).
   On the other hand, good agreement is observed between the unextrapolated H1 data and CASCADE, 
Figure ~\ref{f:f2c} (c).    
\item Statements about the shape of $\ftwoccb_{meas}$ are model dependent:
 \ftwoccb\ rises steeply as $x_{Bj}$ decreases and, when plotted vs. $Q^2$ at constant $x_{Bj}$ ~\cite{zds9697}, does not scale in the measured region. 
This  can be interpreted as coming   from  the rise of the gluon via the PGF dominance of open
charm electroproduction.  \ftwoccb\ represents $\sim 25{\rm -}30$\% of \ftwo\ for
 \qsq$>11$~GeV$^2$ and low $x_{Bj}$. Figure~\ref{f:f2c} (b) has two regimes: it  flattens 
 at low $x_{Bj}$, where both \ftwoccb\ and \ftwo\ are dominated by PGF and, neglecting all mass effects,  $\ftwoccb/ \ftwo\sim \frac{q_c^2}{q_u^2+q_d^2+q_s^2+q_c^2+q_b^2}\sim 4/11\sim 0.36$. 
The denominator of the ratio, \ftwo,  rises at high $x_{Bj}$ due to the valence contribution, forcing the ratio to drop.
\item  Comparing the shape of $\ftwoccb_{meas}$ with other models, or two $\ftwoccb_{meas}$
 extracted assuming different theories is, to a large extent,
 a comparison between two models, rather than with data.   Of this kind are all 
comparisons that can be made in Figure~\ref{f:f2c} (d):
 $\ftwoccb_{meas}$ from three independent samples extracted  
with  HVQDIS (GRV 98, GRV 94, ZEUS NLO) seem to be compatible.
 The prediction using the 
H1 fit agrees reasonably well with the shape of HVQDIS (GRV 98, GRV 94, ZEUS NLO).
\end{enumerate}

\vspace{-0.3cm}

\section{Summary}

\vspace{-0.2cm}

 The overall picture is that of reasonable agreement among all  data samples,
 obscured by the different interpretation of these data, especially concerning the level 
 of  description by  FO NLO pQCD calculations.

 In PHP, H1 observes an adequate description  by FMNR and uses 
these data  to extract 
$x_g g(x_g)$. However, ZEUS observes a deficit in the FMNR prediction, specially 
in the forward region. Alternative models/explanations are massless calculations, BKL, 
fragmentation effects or large NNLO corrections.

The agreement within theoretical uncertainties of the FO NLO pQCD  HVQDIS predictions 
using pdf extracted from DGLAP fits to the inclusive \ftwo\ scaling violations with
the H1 and ZEUS \ds\ production cross section and the ZEUS semileptonic cross section
 provides support for the universality of the parton distributions of the $p$ and 
the validity of factorization in charm  electroproduction. 
 However, H1 finds better agreement with CASCADE predictions than with those of HVQDIS. 

\vspace{-0.3cm}

\section{Outlook}

\vspace{-0.2cm}

 Measurements of open charm production will benefit from the  
expected increase of  a factor of 5 in luminosity,  
together with the instrumental improvements of the H1 and ZEUS detectors 
in HERA Run II. This can result in a 
extended coverage of the phase space, especially in the forward region, 
which  could be used to:
\begin{itemize}
\item Improve the understanding of charm photoproduction.
\item Increment the sensitivity to  IC in DIS, since forward $\eta$ correspond to higher $x_{Bj}$.
\item Look for indications to distinguish between the  VFNS approach and FO NLO.
\item Minimize the extrapolation dependence of \ftwoccb. 
In this case it would be useful to explore channels beyond the usual 
\ds\ modes, whose lower \ptr\ limit is fixed  because of the good 
correlation between \ptds\ and  \ptr($\pi_s$).
\end{itemize}
 Ongoing studies on c-meson spectroscopy~\cite{spectroscopy,zeusdzero,ds1budapest}, diffraction~\cite{dsdiff} or double tagged 
final states will benefit from the increase in statistics. 

 On the theoretical side,  an improved treatment of fragmentation, matching
 the NLO calculation to MC fragmentation models as JETSET, could
lead to  a better description  of the data. It would be also desirable to 
identify a more distinctive signature for CCFM evolution.
 The final goal  is to include open charm cross section data in global pdf fits.

\vspace{-0.3cm}
\section*{Acknowledgements} 
\vspace{-0.2cm}

I thank the organizers of the Ringberg Workshop 
for the nice atmosphere. 
The author  is financially  supported by the Comunidad Aut\'onoma de Madrid.


\vspace{-0.3cm}

\section*{References}
\vspace{-0.2cm}
\begin{thebibliography}{29}

\bibitem{spectroscopy}  Breitweg J \etal [ZEUS Collaboration] 2000, 
Contributed  paper to XXXth ICHEP, abstract 854.

\bibitem{h194} Adloff C \etal [H1 Collaboration] 1996, \ZP~C{\bf 72}, 593.

\bibitem{h196} Adloff C \etal [H1 Collaboration] 1999, \NP B {\bf 545}, 21. 

\bibitem{h197} Adloff C \etal [H1 Collaboration]  2000, 
Contributed  paper to XXXth ICHEP, abstract 984. Mohrdieck S [H1 Collaboration],
 talk at DIS2001, http://dis2001.bo.infn.it/wg/sfwg.html.

\bibitem{dsdijet}  Breitweg J \etal [ZEUS Collaboration] 1999, {\it Eur. Phys. J.} C {\bf 6}, 67.

\bibitem{zds44} Breitweg J \etal [ZEUS Collaboration] 1999, 
Contributed  paper to XXIXth EPS, abstract 525.

\bibitem{zds9697} Breitweg J \etal [ZEUS Collaboration] 2000, {\it Eur. Phys. J.} C {\bf 12}  1, 35.

\bibitem{ds318}  Breitweg J \etal [ZEUS Collaboration] 2000, 
Contributed  paper to XXXth ICHEP, abstract 855.

\bibitem{dsubs} Breitweg J \etal [ZEUS Collaboration] 2000, \PL B {\bf 481 2-4}, 213. 


\bibitem{zeusdzero}
  Chekanov S \etal [ZEUS Collaboration] 2001, 
Contributed  paper to XXXth EPS, abstract 501.

\bibitem{semileptonic} Breitweg J \etal [ZEUS Collaboration] 2000, 
Contributed  paper to XXXth ICHEP, abstract 853.

\bibitem{FRIXIONE}  Frixione S, Nason P and  Ridolfi G  1995, \NP B {\bf 454},  3.

\bibitem{massless} Kniehl B A \etal 1997, \ZP C  {\bf 76}, 689; Binnewies J \etal 1997,
 \ZP C  {\bf 76}, 677; Cacciari M \etal 1997, \PL D {\bf 55}, 2736. 

\bibitem{peterson}
 Peterson C \etal 1983, \PR~D{\bf 27}, 105.

\bibitem{GLADILIN} Gladilin L, hep-ex/9912064.



\bibitem{epsilon0.035} Nason P and Oleari C 2000, \NP B {\bf 565}, 245.

\bibitem{ds1budapest}  
 Chekanov S \etal [ZEUS Collaboration] 2001, 
Contributed  paper to XXXth EPS, abstract 497.


\bibitem{frixionedis01} Frixione S 2001, talk at DIS2001, http://dis2001.bo.infn.it/wg/hiwg.html.


\bibitem{bkl} A.V.Berezhnoy, V.V.Kiselev and  A.K.Likhoded, hep-ph/990555 



\bibitem{HVQDIS} Harris B W and Smith J 1998 , \PR  D {\bf 57} (1998), 2806;
Laenen E, Riemersma S, Smith J and Van~Neerven W L 1995, \NP B{\bf 392},162.


\bibitem{chuvakin} Chuvakin A , Smith J and   Harris B W 2001, {\it Eur. Phys. J.} C {\bf 18}, 547.

\bibitem{CASCADE} Jung H 1999, HERAMC Workshop DESY-PROC-1999-02, 75 


\bibitem{dsdiff} Breitweg J \etal [ZEUS Collaboration] 2000, 
Contributed  paper to XXXth ICHEP, abstract 874.


\end{thebibliography}
\end{document}